\documentstyle[amssymb,prb,aps,multicol,eqsecnum]{revtex}

\renewcommand{\narrowtext}{\begin{multicols}{2}}
\renewcommand{\widetext}{\end{multicols}}

\begin{document}
\title{Non-linear $\sigma $-model for long range disorder and quantum chaos.}
\author{V. R. Kogan$^{1,2}$ and K. B. Efetov${}^{1,2}$}
\address{${}^1$ Theoretische Physik III, \\
Ruhr-Universit\"at Bochum, 44780 Bochum, Germany\\
${}^2$ L. D. Landau Institute for Theoretical Physics, 117940 Moscow, Russia}
\date{\today}
\maketitle
\draft

\begin{abstract}
We suggest a new scheme of derivation of a non-linear ballistic $\sigma $%
-model for a long range disorder and quantum billiards. The derivation is
based on writing equations for quasiclassical Green functions for a fixed
long range potential and exact represention of their solutions in terms of
functional integrals over supermatrices $Q$ with the constraint $Q^{2}=1$.
Averaging over the long range disorder or energy we are able to write a
ballistic $\sigma $-model for all distances exceeding the electron
wavelength (Eq. \ref{e22}). Neither singling out slow modes nor a
saddle-point approximation are used in the derivation. Carrying out a course
graining procedure that allows us to get rid off scales in the Lapunov
region we come to a reduced $\sigma $-model containing a conventional
collision term. For quantum billiards, we demonstrate that, at not very low
frequencies, one can reduce the $\sigma $-model to a one-dimensional $\sigma 
$-model on periodic orbits. Solving the latter model, first approximately
and then exactly, we resolve the problem of repetitions.
\end{abstract}


\bigskip \narrowtext

\section{Introduction}

The $\sigma $-model approach to disordered systems, first written within the
replica trick \cite{wegner,elk}, proved to be a powerful method of
calculations when formulated in the supersymmetric form \cite{efetov}. This
method allows to describe the electron motion at large distances assuming
that at shorter distances the motion is diffusive. For study of such
phenomena as localization, level statistics in a system of a size $L$ much
exceeding the elastic mean free path, etc., the information obtained from
the $\sigma $-model is sufficient. Although the $\sigma $-model is not valid
at distances smaller than the mean free path $l$, the information about the
motion at distances below $l$ is not very interesting for these phenomena.

Success in nano-fabrication made possible producing and studying clean
systems with the size smaller than the elastic mean free path $l$ (for a
review, see. e.g. \cite{kouw}). In such small systems called now quantum
dots electrons move ballistically being scattered mainly the walls. There
are many interesting questions about transport in the quantum dots and
related systems that cannot be answered using the picture of the diffusive
motion. In order to describe the ballistic motion one has to go beyond the
diagrammatic and field theoretical methods developed for disordered systems.

Another motivation to study the ballistic motion originates from the field
called now quantum chaos. The subject of research in the quantum chaos is to
understand the quantum behavior within models that are chaotic in the
classical limit. There are many books and reviews related to this field
(see, e.g. \cite{gutzbook,haake,gian}). The most popular analytical tool for
studying the quantum chaos is the Gutzwiller trace formula \cite{gutz} that
reduces calculation of the density of states to a sum over periodic orbits.
This method (complemented by different approximation schemes) allows one to
study very well the limit of not very long times $t$ when the motion along a
periodic orbit is well defined. At the same time, calculations with the
trace formulae become very difficult in the limit $t\rightarrow \infty $
when one expects an universal behavior described by the Wigner-Dyson
statistics \cite{mehta}.

Both the experimental and theoretical interest to investigations of the
ballistic motion resulted in several attempts to construct a generalization
of the supersymmetric $\sigma $-model to distances smaller than the mean
free path $l$ due to scattering on impurities or the walls in the system.
Muzykantskii and Khmelnitskii (MK) \cite{mk}decoupled as usual the $\psi
^{4} $ interaction in the effective Lagrangian by gaussian integration over
a supermatrix $Q$ but did not use after that a saddle-point approximation.
Instead, they derived a quasi-classical equation for effective Green
functions $g$ analogous to the Eilenberger equation \cite{eilen} well known
in the superconductivity theory. Using an analogy of this equation with an
equation for motion of a magnetic moment in an external magnetic field MK
noticed that the equation was a minimum of a functional $\Phi $ containing a
Wess-Zumino-Novikov-Witten (WZNW) term. So, they replaced the solution of
the semiclassical equations by a functional integral containing the
functional $\Phi $, which allowed to average over the supermatrix $Q$. This
could be done provided the equation of the Green functions corresponded to a
deep minimum of the functional $\Phi $, such that fluctuations near the
minimum could be neglected. Although the authors of Ref. \cite{mk}
conjectured that their field theory could be applicable even in the limit of
a vanishing disorder, they did not confirm this point of view by any
calculations.

A more traditional way of derivation was used by Andreev {\it et al }\cite
{asaa} who tried to derive the $\sigma $-model for a ballistic quantum
billiard. Instead of averaging over disorder they averaged over the energy.
After decoupling the $\psi ^{4}$ term by integration over the supermatrix $%
\tilde{Q}$ they used the saddle-point approximation, which is equivalent to
the self-consistent Born approximation (SCBA). This had to fix the
eigenvalues of $\tilde{Q}$ such that one could put $\tilde{Q}^{2}=1$ and
what remained to do was to expand the action in gradients of $\tilde{Q}$ and
in the frequency $\omega $. Shortly after it became clear that the
saddle-point did not fix the eigenvalues of the supermatrix $\tilde{Q}$ and
modes that were usually massive for disordered system became massless in the
ballistic limit. This problem was discussed in the publications \cite
{zirn,aos}. It is relevant to notice that the existence of the additional
massless modes is not just a consequence of the bad saddle-point
approximation. As discussed in Ref. \cite{zirn} the same problem is
encountered when deriving the $\sigma $-model with the help of the so-called
``color-flavor'' transformation. Within this approach, although one does not
need to use any saddle-point approximation, the expansion in gradients still
remains to be performed. However, there is no parameter that would allow one
to take into account the lowest gradients only. Zirnbauer suggested \cite
{zirn} to perform an additional averaging over ensembles in order to
suppress short range fluctuations.

The saddle-point approximation can actually be useful if one considers a
system with a long range disorder \cite{te1}. In this case, the
single-particle mean free path $l$ can be much smaller than the transport
mean free path $l_{tr}$. At distances exceeding the length $l$ the
saddle-point approximation and the expansion in gradients of $\tilde{Q}$ can
be used and one comes to a ballistic $\sigma $-model that reduces to the
diffusion one only at distances exceeding $l_{tr\text{. }}$ Thus, in the
interval between $l$ and $l_{tr}$ one can obtain the ballistic $\sigma $%
-model in a reliable way (see also a subsequent discussion in Ref. \cite{bmm}%
). However, this does not solve the problem completely because a reasonable
semi-classics should be applicable at all distances exceeding the wavelength 
$\lambda _{F}$.

A ballistic $\sigma $-model should describe low lying excitations that exist
for any long range disorder. At the same time, the conventional saddle-point
approximation, being equivalent to the SCBA, can be good for a short range
disorder only and, hence, may not be used for derivation of a $\sigma $%
-model for a long range disorder and quantum chaos at arbitrary distances.
The same is true for the decoupling of the $\psi ^{4}$ term. The integration
over the supermatrix $\tilde{Q}$ is usually used after singling out slowly
varying pairs $\psi \bar{\psi}$. However, if the random potential is very
long ranged or one averages over the energy, one has slowly varying pairs
from the beginning and there is no necessity of integration over the
supermatrix $\tilde{Q}$ instead of integration over the initial random
potential $U\left( r\right) $. The same is true when applying the
color-flavor transformation \cite{zirn}. The replacement of an integration
over $u\left( r\right) $ by an integration over a supermatrix $Z\left(
r\right) $ does not seem to correspond to physical processes and is a purely
(although exact) mathematical transformation. Thus, the correct scheme of
the derivation of a field theory describing the low lying excitations should
not be based on the Hubbard-Stratonovich or color-flavor decoupling and the
saddle-point approximation.

In this paper, we present a derivation of a ballistic $\sigma $-model using
neither the Hubbard-Stratonovich decoupling with a supermatrix $\tilde{Q}$
nor the saddle-point approximation determining the eigenvalues of $\tilde{Q}$%
. As in Ref. \cite{mk} we derive quasiclassical equations for Green
functions but we write them without making the Hubbard-Stratonovich
transformation. This is justified because \ we use a long range potential.
Only if a short range potential is added we must single out slow pairs in
corresponding term in the Lagrangian and decouple it by the integration over
supermatrices. The crucial step of the derivation is an exact representation
of the solution of the quasiclassical equation in terms of a functional
integral over $8\times 8$ supermatrices $Q_{{\bf n}}\left( {\bf r}\right) $
with the constraint $Q_{{\bf n}}^{2}\left( {\bf r}\right) =1$, where ${\bf r}
$ is the coordinate and ${\bf n=p}_{F}{\bf /}\left| {\bf p}_{F}\right| $ is
the normalized vector on the Fermi-surface. An effective action $\Phi _{u}%
\left[ g_{{\bf n}}\right] $ entering the functional integral is similar the
one written in Ref. \cite{mk}. We show that supersymmetric properties of the
matrix $Q_{{\bf n}}\left( {\bf r}\right) $ make the representation exact,
which was not noticed in the MK variational approach. Moreover, the solution
written for an arbitrary long range potential $u\left( {\bf r}\right) $ is
applicable even for non-averaged quantities.

Averaging over the random potential leads to an effective action $\Phi \left[
g_{{\bf n}}\right] $ that has a form different from those discussed
previously. Analyzing properties of the new ballistic non-linear $\sigma $%
-model with the action $\Phi \left[ g_{{\bf n}}\right] $ we demonstrate that
a new length $l_{L}=v_{F}\tau _{L}$ introduced by Aleiner and Larkin \cite
{al}, where $v_{F}$ is the Fermi velocity and $\tau _{L}$ is the inverse
Lapunov exponent, determines different regimes. The importance of this
length was also discussed recently in Ref. \cite{gm}.\ Integrating over
variations of the supermatrix $Q_{{\bf n}}\left( {\bf r}\right) $ at
distances smaller than $l_{L}$ we come to another form of the ballistic $%
\sigma $-model containing the conventional collision term.

We show that without an internal disorder the calculation of the functional
integral can be reduced to study of the $\sigma $-model for periodic orbits.
Only the presence of a regularizer analogous to the one introduced in Ref. 
\cite{al} may mix the periodic orbits. The problem of repetitions \cite{bk}
is discussed and we are able to demonstrate that the contradiction between
the references \cite{bk} and \cite{asaa} is rather a consequence of an
unjustified approximation used in Ref. \cite{asaa} than a deficiency of the $%
\sigma $-model.

The paper is organized as follows: In Chapter II, we express correlation
functions of interest in terms of functional integrals over supervectors and
write equations for generalized Green functions. In Chapter III, we
represent the solution of the quasiclassical equations in terms of
functional integrals over supermatrices and average over disorder, thus
obtaining a ballistic $\sigma $-model applicable at all distances exceeding
the wavelength. In Chapter IV, we integrate over a Lapunov region and derive
a reduced ballistic $\sigma $-model containing a collision term. In Chapter
V, we show how one can derive equations for correlation functions. In
Chapter VI, we show how calculations within the ballistic $\sigma $-model
can be reduced to calculations for periodic orbits. We explain how the so
called ``repetition problem'' can be resolved. Chapter VII is devoted to a
discussion of the results obtained. The Appendix contains a derivation of
the boundary conditions.

\section{Formulation of the problem. Quasiclassical approximation.}

The aim of the present paper is to find a convenient representation that
would allow us to consider an electron motion in a smooth potential at large
times or low frequencies. Of course, with the formalism presented one can
consider wave scattering in microwave cavities and other interesting
problems but, to simplify notations, we will use the condensed matter
language.

We want to extend the supersymmetry method \cite{efetov} developed for
disordered systems to distances smaller than the mean free path. Actually,
the only assumption we will use in the derivation is that all physical
quantities vary at distances exceeding the Fermi wavelength $\lambda
_{F}=2\pi p_{F}^{-1.}$. Although this assumption is much less restrictive
than those used in Ref. \cite{efetov} it allows to simplify essentially the
consideration. Averaging over the energy, which is the standard procedure
for quantum chaos, can be considered as the limiting case for an infinite
range random potential.

The method developed in this paper is applicable for calculation of gauge
invariant quantities like density-density or level-level correlation
functions. Such quantities as average one-particle Green functions at
different points will not be considered here. We choose the Hamiltonian ${%
\hat{H}}$ of the system in the standard form 
\begin{equation}
{\hat{H}}={\hat{H}}_{0}+u\left( {\bf r}\right) +u_{s}\left( {\bf r}\right) ,
\label{e1}
\end{equation}

\[
\hat{H}_{0}=-{\bf \nabla }^{2}/2m-\varepsilon _{F} 
\]
where $u\left( {\bf r}\right) $ is a long range potential, which is of the
main interest now, and $u_{s}\left( {\bf r}\right) $ is a short range
impurity potential. The latter is added in order to make the models somewhat
more general. The presence of the short range potential will help to
understand better the procedure we will use. However, nothing is assumed
about the strength of $u_{s}\left( {\bf r}\right) $ and it can be safely put
to zero in all formulae written below.

As usual \cite{efetov}, one can express correlation functions of interest in
terms of a functional integral over $8$-component supervectors $\psi \left( 
{\bf r}\right) $ with an effective Lagrangian $L$%
\[
L\left[ \psi \right] =\int [-i\bar{\psi}\left( {\bf r}\right) \left( \hat{%
\tilde{H}}_{0}+u\left( {\bf r}\right) \right) \psi \left( {\bf r}\right) 
\]
\begin{equation}
-\frac{i\left( \omega +i\delta \right) }{2}\bar{\psi}\left( {\bf r}\right)
\Lambda \psi \left( {\bf r}\right) +\frac{1}{4\pi \nu \tau _{s}}\left( \bar{%
\psi}\left( {\bf r}\right) \psi \left( {\bf r}\right) \right) ^{2}]d{\bf r,}
\label{e2}
\end{equation}
\[
\hat{\tilde{H}}_{0}=\hat{H}_{0}-\varepsilon +\frac{\omega }{2} 
\]
where $\varepsilon $ is the energy at which the physical quantities are
considered and $\omega $ is the frequency.

The Lagrangian $L$, Eq. (\ref{e2}), is written after the averaging over the
short range potential $u_{s}\left( {\bf r}\right) $ implying the standard
gaussian correlations of the type 
\begin{equation}
\langle u_{s}\left( {\bf r}\right) u_{s}\left( {\bf r}^{\prime
}\right)\rangle =\frac{1}{2\pi \nu \tau _{s}}\delta \left( {\bf r-r}^{\prime
}\right)  \label{e3}
\end{equation}

It is relevant to emphasize that averaging over the long range potential $%
u\left( {\bf r}\right) $ has not been performed.

The correlation functions we are interested in can be obtained adding proper
source terms in the Lagrangian $L\left[ \psi \right] $. An important class
of the correlation functions can be obtained writing the Lagrangian $L_{a}%
\left[ \psi \right] $ including the sources in the form 
\begin{equation}
L_{a}\left[ \psi \right] =L\left[ \psi \right] +i\int \bar{\psi}\left( {\bf r%
}\right) \hat{a}\left( {\bf r}\right) \psi \left( {\bf r}\right) d{\bf r}
\label{e4}
\end{equation}
where $\hat{a}\left( {\bf r}\right) $ is a matrix depending on coordinates.
Its explicit form depends on what type of the correlation function is
calculated.

The level-level correlation function $R\left( \omega \right) $ 
\[
R\left( \omega \right) =\frac{1}{2\pi ^{2}\omega \nu ^{2}V^{2}}\langle 
\mathop{\rm Re}%
\int \left( n\left( \varepsilon -\omega \right) -n\left( \varepsilon \right)
\right) 
\]
\begin{equation}
\times G_{\varepsilon -\omega }^{A}\left( {\bf r,r}\right) \left(
G_{\varepsilon }^{R}\left( {\bf r}^{\prime }{\bf ,r}^{\prime }\right)
-G_{\varepsilon }^{A}\left( {\bf r}^{\prime },{\bf r}^{\prime }\right)
\right) d{\bf r}d{\bf r}^{\prime }d\varepsilon\rangle,  \label{e5}
\end{equation}
where $n\left( \varepsilon \right) $ is the Fermi distribution, contains the
product $G_{\varepsilon -\omega }^{A}G_{\varepsilon }^{R}$. For calculation
of this product one should choose $\hat{a}$ in the form 
\begin{equation}
\hat{a}=\left( 
\begin{array}{cc}
\hat{\alpha}_{1} & 0 \\ 
0 & -\hat{\alpha}_{2}
\end{array}
\right) ,\text{ \ \ }\hat{\alpha}_{1,2}=\frac{\alpha _{1,2}}{2}\left(
1-k\right)  \label{e6}
\end{equation}
where both $k$ and $\tau _{3}$ denoting different blocks have the form 
\[
\left( 
\begin{array}{cc}
1 & 0 \\ 
0 & -1
\end{array}
\right) 
\]
Below we use the same notations as in the book \cite{efetov}.

The product $\langle G_{\varepsilon -\omega }^{A}G_{\varepsilon }^{A}\rangle$
is trivial, whereas the first product in Eq. (\ref{e5}) can be written as 
\[
\langle G_{\varepsilon -\omega }^{A}\left( {\bf r,r}\right) G_{\varepsilon
}^{R}\left( {\bf r}^{\prime },{\bf r}^{\prime }\right)\rangle 
\]
\[
=-4\int \psi _{4}^{1}\left( {\bf r}\right) \bar{\psi}_{4}^{1}\left( {\bf r}%
\right) \psi _{4}^{2}\left( {\bf r}^{\prime }\right) \bar{\psi}%
_{4}^{2}\left( {\bf r}^{\prime }\right) \exp \left( -L_{a}\left[ \psi \right]
\right) D\psi 
\]
and we reduce the function $R\left( \omega \right) $ to the form 
\begin{equation}
R\left( \omega \right) =\frac{1}{2}  \label{e7}
\end{equation}
\[
-\frac{1}{2\left( \pi \nu V\right) ^{2}}\lim_{\alpha _{1}=\alpha _{2}=0}%
\mathop{\rm Re}%
\int \frac{\partial ^{2}}{\partial \alpha _{1}\partial \alpha _{2}}\exp
\left( -L_{a}\left[ \psi \right] \right) D\psi 
\]
For computation of the density-density correlation function one has to
calculate the averages of the type 
\begin{equation}
Y^{00}({\bf r}_{1}{\bf ,r}_{2};\omega )=2\langle G_{\varepsilon -\omega
}^{A}\left( {\bf r}_{2}{\bf ,r}_{1}\right) G_{\varepsilon }^{R}\left( {\bf r}%
_{1},{\bf r}_{2}\right)\rangle  \label{e7a}
\end{equation}

\bigskip

This product can be obtained from the following source term 
\begin{equation}
\hat{a}=\left( 
\begin{array}{cc}
0 & \hat{\alpha}_{0}\left( {\bf r}\right) \\ 
-\hat{\alpha}_{0}\left( {\bf r}\right) & 0
\end{array}
\right)
\end{equation}
where 
\[
\hat{\alpha}_{0}\left( {\bf r}\right) =\left( 
\begin{array}{cc}
\hat{\alpha}_{1}\delta \left( {\bf r-r}_{1}\right) & 0 \\ 
0 & \hat{\alpha}_{2}\delta\left( {\bf r-r}_{2}\right)
\end{array}
\right) 
\]
with the same expression for $\hat{\alpha}_{1,2}$ as in Eq. (\ref{e6}).
Then, we have for $Y^{00}\left( {\bf r}_{1}{\bf ,r}_{2},\omega \right) $%
\begin{equation}
Y^{00}\left( {\bf r}_{1},{\bf r}_{2};\omega \right) =2\int \left. \frac{%
\partial ^{2}}{\partial \alpha _{1}\partial \alpha _{2}}\exp \left( -L_{a}%
\left[ \psi \right] \right) D\psi \right| _{\alpha _{1}=\alpha _{2}=0}
\label{e9}
\end{equation}

In principle we can proceed calculating in Eqs. (\ref{e7}, \ref{e9}) in the standard way 
\cite{efetov} by singling out slowly varying pairs $\psi \bar{\psi}$ in the
term $\psi ^{4}$ in the Lagrangian $L$, Eqs. (\ref{e2}, \ref{e4}), and
decoupling the products of these terms by Gaussian integration over $8\times
8$ supermatrices $M$. A word of caution should be said at this point. The
separation into the products of slowly varying pairs $\psi \bar{\psi}$ makes
a sense only for distances exceeding the range of the random potential. This
means that one may not this approximation for distances smaller than the
potential range. Of course, the same is true when averaging over the
spectrum. In the latter case the range of the random potential is just the
system size. Therefore, the previous derivations where this separation was
used \cite{asaa} can hardly be justified. The same problem apparently arises
when doing the color-flavor transformation \cite{zirn}. Although the
transformation is formally exact, there is no reason for neglecting higher
gradients when making the expansions in gradients.

In order to avoid the problem we keep the long range potential $u\left( {\bf %
r}\right) $ in Eq. (\ref{e2},\ref{e4}) as it stands and do not average over
it. This will be done later. At the same time, the decoupling of the term
corresponding to the short range potential by integration over the
supermatrix $M$ can be safely done (We repeat again that the presence of the
short range potential is not crucial for our derivation and its strength can
be put zero).

After the decoupling for the short range potential the calculation of the
correlation functions is reduced to the computation of an effective
partition function $Z_{1}\left[ J\right] $%
\begin{equation}
Z_{1}\left[ J\right] =\int \exp \left( -L_{J}\left[ \psi \right] \right)
D\psi  \label{e10}
\end{equation}
where 
\begin{equation}
J\left( {\bf r}\right) =i\hat{a}\left( {\bf r}\right) +\frac{M\left( {\bf r}%
\right) }{2\tau _{s}}  \label{e11}
\end{equation}
and $L_{J}[\psi ]$ is obtained from $L_{a}[\psi ]$, Eqs. (\ref{e2}), (\ref
{e4}), after the replacement of the parameter $\hat{a}$ by $J$ according to
Eq. (\ref{e11}) and neglecting the quartic term. The function $J$ satisfies
the standard symmetry relation $J=\bar{J}$, where $\bar{J}=CJ^{T}C^{T}$. The
bar means the usual ``charge conjugation'' of Refs. \cite{efetov}, the
matrix $C$ is defined as follows 
\[
C=\Lambda \otimes \left( 
\begin{array}{cc}
c_{1} & 0 \\ 
0 & c_{2}
\end{array}
\right) ,\text{ \ }c_{1}=\left( 
\begin{array}{cc}
0 & -1 \\ 
1 & 0
\end{array}
\right) ,\text{ \ }c_{2}=\left( 
\begin{array}{cc}
0 & 1 \\ 
1 & 0
\end{array}
\right) 
\]

The correlation functions for a given long range potential $u\left( {\bf r}%
\right) $ can be calculated by differentiating in $\alpha _{1,2}$ the
following integral 
\begin{equation}
Z_{u}=\int Z_{1}\left[ J\right] \exp \left( -\frac{\pi \nu }{8\tau _{s}}\int
StrM^{2}\left( {\bf r}\right) d{\bf r}\right) DM  \label{e12}
\end{equation}
Averaged correlation functions can be obtained from the quantity $Z=$ $%
\langle Z\rangle_{u}$, where $\langle\dots\rangle_{u}$ means averaging over $%
u\left( {\bf r}\right) $. Of course, we could immediately average over $%
u\left( {\bf r}\right) $ in Eq. (\ref{e12}) but we want to avoid the
standard scheme. The approximations that worked so well for short range
potential are not applicable to the long range one. In particular, not only
the separation of slow modes is not justified but also the saddle-point
approximation for the integral over supermatrices is no longer good. It is
clear that the existence of low lying excitations like diffusons and
cooperons is more general than the SCBA and one should try to avoid the
latter.

In this paper we follow the method of quasiclassical Green functions first
introduced for study of superconductivity \cite{eilen}. This method was used
recently by Muzykantskii and Khmelnitskii \cite{mk} for study of the
ballistic transport. Our calculations are partially equivalent to those by
MK \cite{mk} but there are essential differences. First, we keep the long
range potential $u\left( {\bf r}\right) $ fixed and average over it at the
later stage of the derivation. Second, which is the most crucial step, we
show how to write the solution of the quasiclassical equations ${\em exactly}%
{\it .}$ The possibility of writing an exact solution of the equations for
quasiclassical Green functions in terms of functional integrals is a
consequence of the supersymmetry and has not been realized before.

As in MK \cite{mk}, we consider the logarithmic derivative of the partition
function $Z_{1}\left[ J\left( {\bf r}\right) \right] $ 
\begin{equation}
\frac{\delta \ln Z_{1}[J({\bf r})]}{\delta J_{\alpha \beta }({\bf r})}%
=K_{\beta \beta }\langle \psi _{\alpha }({\bf r}){\bar{\psi}}_{\beta }({\bf r%
}^{\prime })\rangle _{\psi }  \label{e13}
\end{equation}
where $<...>_{\psi }$ is the average with the functional $L_J\left[ \psi %
\right] ,$ Eqs. (\ref{e2}, \ref{e4}, \ref{e11}) and 
\[
K=\left( 
\begin{array}{cc}
k & 0 \\ 
0 & k
\end{array}
\right) 
\]
Introducing the matrix function $G\left( {\bf r,r}^{\prime }\right) $ as

\begin{equation}
G({\bf r},{\bf r}{^{\prime }})=2\langle \psi ({\bf r}){\bar{\psi}}({\bf r}{%
^{\prime }})\rangle _{\psi }  \label{e14}
\end{equation}
we can write for this function the following equation 
\begin{equation}
\left[ {\hat{\tilde H}}_{0{\bf r}}+u\left( {\bf r}\right) +\Lambda \frac{%
\omega +i\delta }{2}+iJ\left( {\bf r}\right) \right] G\left( {\bf r,r}%
^{\prime }\right) =i\delta \left( {\bf r-r}^{\prime }\right)
\label{eqgreenfunc}
\end{equation}
where the subscript ${\bf r}$ of $\hat{\tilde{H}}_{0{\bf r}}$ means that the
operator acts on ${\bf r}$.

Conjugating Eq.(\ref{eqgreenfunc}) and using the property 
\begin{equation}
\bar{G}({\bf r},{\bf r}^{\prime })=CG^{T}({\bf r^{\prime}},{\bf r})C^{T}=
G\left( {\bf r,r}^{\prime }\right)  \label{e14a}
\end{equation}
we obtain another equation for the matrix $G({\bf r},{\bf r}^{\prime })$
with the operator $\tilde{H}_{0{\bf r}^{\prime }}$ acting on its second
variable

\begin{equation}
G({\bf r};{\bf r}{^{\prime }})\left[ \hat{\tilde{H}}_{0{\bf r}^{\prime
}}+u\left( {\bf r}^{\prime }\right) +\Lambda \frac{\omega +i\delta }{2}+iJ(%
{\bf r}{^{\prime }})\right] =i\delta ({\bf r}-{\bf r}{^{\prime }})
\label{conjeqgreenfunc}
\end{equation}

Until now no approximations have been done and Eqs. (\ref{eqgreenfunc}, \ref
{conjeqgreenfunc}) are exact. Now we can use the assumption that the
potential $u\left( {\bf r}\right) $ changes slowly on the wavelength $%
\lambda _{F}$. If the mean free path for the scattering on the random
potential exceeds $\lambda _{F}$ the Green function varies as a function of $%
{\bf r-r}^{\prime }$ at distances of the order of $\lambda _{F}$ but, at the
same time, is a slow function of $\left( {\bf r+r}^{\prime }\right) /2$. The
Fourier transform $G_{{\bf p}}\left( \left( {\bf r+r}^{\prime }\right)
/2\right) $ of $G$ $\left( {\bf r,r}^{\prime }\right) $ respective to ${\bf %
r-r}^{\prime }$ has a sharp maximum near the Fermi surface. In order to
cancel large terms we subtract Eq. (\ref{conjeqgreenfunc}) from Eq. (\ref
{eqgreenfunc}). Using the assumption that the potential $u\left( {\bf r}%
\right) $ is smooth and expanding it in gradients we obtain in the lowest order

\begin{equation}
\left[ -\frac{i{\bf p\nabla }_{{\bf R}}}{m}+i{\bf \nabla }_{{\bf R}}u({\bf R}%
)\frac{\partial }{\partial {\bf p}}\right] G_{{\bf p}}({\bf R})  \label{e15}
\end{equation}

\[
+\frac{\omega +i\delta }{2}[\Lambda ,G_{{\bf p}}({\bf R})]+i[J({\bf R}),G_{%
{\bf p}}({\bf R})]=0, 
\]
where \ ${\bf R}=\left( {\bf r}+{\bf r}{^{\prime }}\right) /2$ and $\left[ ,%
\right] $ stands for the commutator.

When deriving Eq. (\ref{e15}), not only the potential $u\left( {\bf r}%
\right) $ but also the function $J\left( {\bf r}\right) $ was assumed to be
smooth. The supermatrix $M\left( {\bf r}\right) $ is smooth by the
construction. As concerns $\hat{a}$, this is not always so, as is seen from
Eq. (\ref{e7a}-\ref{e9}). However, we can slightly smear the coordinates in
the definition of the correlation functions and obtain after this procedure
a smooth function $\hat{a}$.

The dependence of the Green function $G_{{\bf p}}\left( {\bf R}\right) $ on $%
{\bf |p|}$ is more sharp than on other variables. In order to avoid this
sharp dependence we integrate Eq. (\ref{e15}) over $|{\bf p|}$. Of course,
this procedure makes a sense for very large samples when the level
discreteness can be neglected. However, this procedure can also be performed
in finite samples provided it is complemented by an averaging over the
energy.

The most interesting contribution in the integral over $|{\bf p|}$ comes
from the vicinity of the Fermi-surface. A contribution given by momenta
considerably different from $p_{F}$ is proportional to the unity matrix an
drops out from Eq. (\ref{e15}).

Introducing the function $g_{{\bf n}}\left( {\bf r}\right) $ 
\begin{equation}
g_{{\bf n}}\left( {\bf r}\right) =\frac{1}{\pi }\int G_{p{\bf n}}\left( {\bf %
r}\right) d\xi \text{, \ }\xi =\frac{p^{2}-p_{F}^{2}}{2m}  \label{klassaver}
\end{equation}
where ${\bf n}$ is a unite vector pointing a direction on the Fermi surface,
we obtain the final quasiclassical equation

{\large 
\begin{eqnarray}
&&\left( v_{F}{\bf n\nabla }-p_{F}^{-1}{\bf \nabla }_{{\bf r}}u\left( {\bf r}%
\right) {\bf \partial }_{{\bf n}}\right) g_{{\bf n}}\left( {\bf r}\right)
\label{eqklassfunc} \\
&&+\frac{i\left( \omega +i\delta \right) }{2}[\Lambda ,g_{{\bf n}}\left( 
{\bf r}\right) ]-[J,g_{{\bf n}}]=0  \nonumber
\end{eqnarray}
} where 
\[
{\bf \partial}_{{\bf n}}={\bf \nabla}_{{\bf n}}-{\bf n} 
\]
\[
{\nabla }_{{\bf n}}=-[{\bf n\times \lbrack n\times }\frac{\partial }{%
\partial {\bf n}}]] 
\]

The function $g_{{\bf n}}\left( {\bf r}\right) $ is self-conjugate 
\begin{equation}
\bar{g}_{{\bf n}}\left( {\bf r}\right) =Cg_{-{\bf n}}^{T}\left( {\bf r}%
\right) C^{T}=g_{{\bf n}}\left( {\bf r}\right) ,  \label{e15a}
\end{equation}
which follows from Eqs. (\ref{e14a}), (\ref{klassaver}).

Eq. (\ref{eqklassfunc}) should be complemented by a boundary condition at
the surface of the sample. This boundary condition is derived in Appendix A.
Considering a closed sample we assume that the current across the border is
equal to zero. This leads the boundary condition at the surface 
\begin{equation}
\left. g_{{\bf n}_{\perp }}\left( {\bf r}\right) \right| _{surface}=\left.
g_{-{\bf n}_{\perp }}\left( {\bf r}\right) \right| _{surface}  \label{e15b}
\end{equation}
where ${\bf n}_{\perp }$ is the component of the vector ${\bf n}$
perpendicular to the surface.

The correlation functions considered here can easily be expressed in terms
of the quasiclassical Green functions $g_{{\bf n}}\left( {\bf r}\right) $.
The functional derivative of the partition functions $Z_{1}$, Eq. (\ref{e13}%
), can be written through the function $g_{{\bf n}}\left( {\bf r}\right) $
as follows

{\large 
\begin{eqnarray}
&&\langle \psi _{\alpha }({\bf r}){\bar{\psi}}_{\beta }({\bf r})\rangle
_{\psi }=\frac{1}{2}G_{\alpha \beta }({\bf r},{\bf r})  \nonumber \\
&=&\frac{\nu }{2}\int d{\bf n}\int d\xi G_{p{\bf n,\alpha \beta }}({\bf r}%
)\approx \frac{\pi \nu }{2}\int d{\bf n}g_{{\bf n,\alpha \beta }}({\bf r})
\label{klassaverage}
\end{eqnarray}
}

In Eq. (\ref{klassaverage}) and everywhere below the symbol $\int d{\bf n}$
implies integration over the unit $d$-dimensional sphere normalized to its
surface area $S_{d}$. With this definition we have e.g. 
\begin{equation}
\int d{\bf n}=1  \label{e15c}
\end{equation}

As in the theory of superconductivity, the solution for the Eq.(\ref
{eqklassfunc}) satisfies the condition $g_{{\bf n}}^{2}({\bf r})=\hat{1}.$

Eq. (\ref{eqklassfunc}) is written for a non-averaged potential $u\left( 
{\bf r}\right) $. It is valid also in the absence of the long range
potential. If the system is finite, Eq. (\ref{eqklassfunc}) can still be
obtained provided averaging over the spectrum is performed.

In principle, one could average over the potential $u\left( {\bf r}\right) $
by expanding in this potential and averaging terms of the perturbation
theory. This would not be considerably more convenient as compared to the
conventional perturbation theory. Fortunately, exact solutions of Eq. (\ref
{eqklassfunc}) can be written explicitly in terms of a functional integral
over supermatrices $Q_{{\bf n}}\left( {\bf r}\right) $ with the same
structure as the supermatrices $g_{{\bf n}}\left( {\bf r}\right) $.

In the next chapter we write this solution and average over the long range
potential $u\left( {\bf r}\right) .$

\section{Solution of the quasiclassical equation. The ballistic $\protect%
\sigma $-model.}

The quasiclassical equation, Eq. (\ref{eqklassfunc}), has been written
previously by MK, Ref. \cite{mk} (the long range random potential $u\left( 
{\bf r}\right) $ and the source term were not included). However, they did
not try to solve this equation but noticed in analogy with an equation of
motion for a ferromagnet that Eq. (\ref{eqklassfunc}) is an extremum of a
functional containing a Wess-Zumino-Novikov-Witten term. Assuming that this
had to be a deep minimum MK have written the solution of the quasiclassical
equation in terms of a functional integral with this functional.
Unfortunately, conditions providing such a description have not been found
as yet and the MK theory is usually spoken of as a variational or
phenomenological one. Although this is generally so for ``equations of
motion'' like the quasiclassical Eq. (\ref{eqklassfunc}), there is an
important exception when the functional integral becomes an exact solution
of the problem. This exception is just the case considered here when the
equation is linear in $g_{{\bf n}}({\bf r})$ and the functions $g_{{\bf n}}(%
{\bf r})$ are supermatrices. Surprisingly, this rather simple fact has not
been realized before.

It is clear that a solution for Eq. (\ref{eqklassfunc}) is not unique
because the equation is homogenous. To determine the solution uniquely one
should put $J=0$, $u\left( {\bf r}\right) =0$ and calculate the
corresponding $g_{{\bf n}}^{\left( 0\right) }\left( {\bf r}\right) $
directly without using the quasiclassical equation for the Green function.
It is easy to see that in this case 
\begin{equation}
g_{{\bf n}}^{\left( 0\right) }\left( {\bf r}\right) =\Lambda  \label{e16a}
\end{equation}

Eq. (\ref{e16a}) plays the role of a boundary condition. Let us show that
the exact solution for Eq. (\ref{eqklassfunc}) satisfying the boundary
condition, Eq. (\ref{e16a}) can be written as

\begin{eqnarray}
&&g_{{\bf n}}({\bf r})=Z_{2}^{-1}[J]\int_{Q_{{\bf n}}^{2}=1}Q_{{\bf n}}({\bf %
r})\exp \left( -\frac{\pi \nu }{2}\Phi _{J}[Q_{{\bf n}}({\bf r})]\right) DQ_{%
{\bf n}},  \nonumber \\
&&\Phi _{J}[Q_{{\bf n}}({\bf r})]=Str\int d{\bf r}d{\bf n[}\Lambda {\bar{T}}%
_{{\bf n}}({\bf r})(v_{F}{\bf n\nabla }_{{\bf r}}  \nonumber \\
&&-p_{F}^{-1}{\bf \nabla }_{{\bf r}}u({\bf r}){\bf \nabla }_{{\bf n}})T_{%
{\bf n}}({\bf r})+\left( \frac{i\left( \omega +i\delta \right) }{2}\Lambda
-J({\bf r})\right) Q_{{\bf n}}({\bf r})],  \label{sigmaaverage} \\
&&Q_{{\bf n}}({\bf r})=T_{{\bf n}}({\bf r})\Lambda {\bar{T}}_{{\bf n}}({\bf r%
}),\text{ \ }{\bar{T}}_{{\bf n}}({\bf r})T_{{\bf n}}\left( {\bf r}\right) =1
\nonumber
\end{eqnarray}
In Eq. (\ref{sigmaaverage}), the partition function $Z_{2}[J\left( {\bf r}%
\right) ]$ is 
\begin{equation}
Z_{2}[J]=\int_{Q_{{\bf n}}^{2}=1}\exp \left( -\frac{\pi \nu }{2}\Phi _{J}[Q_{%
{\bf n}}({\bf r})]\right) DQ_{{\bf n}}  \label{integ}
\end{equation}
and the integration is performed over the self-conjugate supermatrices $Q_{%
{\bf n}}=\bar{Q}_{{\bf n}}\left( {\bf r}\right) $ satisfying the following
relations 
\begin{equation}
Q_{{\bf n}}^{2}\left( {\bf r}\right) =1  \label{e16b}
\end{equation}
everywhere in the bulk and, in addition, 
\begin{equation}
\left. Q_{{\bf n}_{\perp }}\left( {\bf r}\right) \right| _{surface}=\left.
Q_{-{\bf n}_{\perp }}\left( {\bf r}\right) \right| _{surface}  \label{e16c}
\end{equation}
at the surface of the sample.

In order to prove the statement that the integral, Eq.(\ref{sigmaaverage}),
is equal to the solution $g_{{\bf n}}({\bf r})$ for Eq.(\ref{klassaver}), we
notice that the integration in Eq. (\ref{integ}) is performed over all
supermatrices $Q_{{\bf n}}\left( {\bf r}\right) $ with the constraints, Eqs$%
. $(\ref{e16b}, \ref{e16c}), and therefore cannot change under the following
replacement of the variable of the integration 
\begin{equation}
Q_{{\bf n}}({\bf r})\rightarrow \widetilde{Q}_{{\bf n}}({\bf r})=U_{{\bf n}}(%
{\bf r})Q_{{\bf n}}({\bf r})\bar{U}_{{\bf n}}\left( {\bf r}\right) \text{, \
\ \ }U_{{\bf n}}({\bf r})\bar{U}_{{\bf n}}\left( {\bf r}\right) =1
\label{e16d}
\end{equation}
\ On the other hand, one can formally consider the integral, Eq.(\ref{integ}%
), with the transformed matrix $\widetilde{Q}_{{\bf n}}({\bf r})$ as a
functional of the transformation matrix $U_{{\bf n}}({\bf r})$. The fact
that the integral $Z_{2}[J\left( {\bf r}\right) ],$ Eq.(\ref{integ}), does
not change under the transformation means, in particular, that the first
variation of the considered functional must be zero for any unitary matrix $%
U_{{\bf n}}({\bf r})$.

In order to find the first variation, we make a small rotation, resulting in
the replacement $U_{{\bf n}}({\bf r})\rightarrow U_{{\bf n}}({\bf r})+\delta
U_{{\bf n}}({\bf r})U_{{\bf n}}({\bf r})$ and compute in the linear
approximation in the matrix $\delta U_{{\bf n}}({\bf r})$ the difference
between the functionals with the changed and initial matrices $U_{{\bf n}}(%
{\bf r})$. For the supermatrix $\widetilde{Q}_{{\bf n}}({\bf r})$, this
means the replacement $\widetilde{Q}_{{\bf n}}({\bf r})\rightarrow 
\widetilde{Q}_{{\bf n}}({\bf r})+[\delta U_{{\bf n}}({\bf r}),\widetilde{Q}_{%
{\bf n}}({\bf r})]$, which follows from the relation $\delta \bar{U}_{{\bf n}%
}\left( {\bf r}\right) =-\delta U_{{\bf n}}\left( {\bf r}\right) $. Then,
the first variation $\delta Z_{2\text{ }}$of the partition function $%
Z_{2}[J\left( {\bf r}\right) ],$ Eq.(\ref{integ}) takes the form :

\begin{equation}
\delta Z_{2}[J]=-\frac{\pi \nu }{2}\int_{Q_{{\bf n}}^{2}=1}DQ_{{\bf n}%
}\delta \Phi \lbrack Q_{{\bf n}}]\exp \left( -\frac{\pi \nu }{2}\Phi \lbrack
Q_{{\bf n}}]\right)  \label{1variation}
\end{equation}
The variation $\delta \Phi \lbrack Q_{{\bf n}}]$ can easily be calculated
from Eq.(\ref{sigmaaverage}) and we write it as

\[
\delta \Phi \lbrack Q_{{\bf n}}]=-Str\int d{\bf r}d{\bf n}\delta U_{{\bf n}}(%
{\bf r})\{(v_{F}{\bf n\nabla } 
\]

\[
-p_{F}^{-1}{\bf \nabla }_{{\bf r}}u\left( {\bf r}\right) {\bf \partial}_{%
{\bf n}})Q_{{\bf n}}({\bf r})+[(i\frac{\omega +i\delta }{2}\Lambda -J\left( 
{\bf r}\right) ),Q_{{\bf n}}\left( {\bf r}\right) ]\} 
\]
\begin{equation}
+v_{F}Str\int_{S}d{\bf n}\left( {\bf n}d{\bf S}\right) \delta U_{{\bf n}%
}\left( {\bf r}\right) Q_{{\bf n}}\left( {\bf r}\right)  \label{e17a}
\end{equation}

The integration in the last term in Eq. (\ref{e17a}) is performed over the
surface of the sample. At the surface, not only $Q_{{\bf n}}\left( {\bf r}%
\right) $ but also $\delta U_{{\bf n}}\left( {\bf r}\right) $ must be
invariant under the replacement ${\bf n}_{\perp }{\bf \rightarrow -n}_{\perp
}$ (see Eq. (\ref{e16c})) It follows from this property that the surface term
in Eq. (\ref{e17a}) is equal to zero. Then, substituting Eq. (\ref{e17a})
into Eq. (\ref{1variation}) and taking into account that $\delta Z_{2}[J]$
must be zero for any $\delta U_{{\bf n}}\left( {\bf r}\right) $ we come
immediately to Eq. (\ref{eqklassfunc}).

Eq.(\ref{eqklassfunc}) is a differential one and a class of different
solutions may exist. The fact that the integral, Eq. (\ref{sigmaaverage}),
satisfies Eq. (\ref{eqklassfunc}), does not guarantee that it is equal to $%
g_{{\bf n}}\left( {\bf r}\right) $, Eq. (\ref{klassaver}), and this should
be checked separately. It is the supersymmetric structure of the
supermatrices $Q_{{\bf n}}\left( {\bf r}\right) $ that allows the integral,
Eq. (\ref{sigmaaverage}), to satisfy both the equations.

Putting $J=0$ and $u\left( {\bf r}\right) =0$ one can calculate $g_{{\bf n}%
}^{\left( 0\right) }\left( {\bf r}\right) $ directly from Eq. (\ref
{klassaver}) and come to Eq. (\ref{e16a}). The same can be done using the
functional integral in Eq. (\ref{sigmaaverage}). As the functional $\Phi
_{J}[Q_{{\bf n}}({\bf r})]$, Eq.(\ref{sigmaaverage}), contains at $J=0$, $%
u\left( {\bf r}\right)=0$ only the matrices $Q_{{\bf n}}\left({\bf r}\right)$
and $\Lambda $, averaging with such a functional gives according to general
rules \cite{efetov} 
\begin{equation}
\langle Q_{{\bf n}}\left( {\bf r}\right)\rangle_{\Phi }=\Lambda  \label{e18}
\end{equation}
Comparing Eqs. (\ref{e18}) with Eq. (\ref{e16a}) we conclude that the
solution, Eq. (\ref{sigmaaverage}), is compatible with Eq. (\ref{klassaver}%
). Although Eq. (\ref{e18}) is trivially fulfilled for the supermatrices, it 
would not be necessarily correct if the matrices $Q$ did not have the 
supersymmetric structure. In the latter case one could obtain e.g. a non-trivial 
function of $\omega $ in the r.h.s. of it. This would invalidate the present 
approach for such symmetries. Actually, Eq. (\ref{e18}) tells us that the 
average density of states is a constant, which is the case for weakly disordered 
systems.

The partition functions $Z_{1}$, Eqs. (\ref{e2}, \ref{e4}, \ref{e10}, \ref
{e11}), and $Z_{2}$, Eq. (\ref{integ}) are equal to unity at $J=0$. As their
logarithmic derivatives coincide for all $J$, we come to the conclusion that 
\begin{equation}
Z_{1}[J]=Z_{2}[J]  \label{e19}
\end{equation}
Thus, we replaced the integration over electron modes, Eqs. (\ref{e2}, \ref
{e4}, \ref{e10}, \ref{e11}), by the integration over low lying excitations
that were called diffusion modes in the diffusive limit. Apparently, the
name kinetic modes would be a more proper one for the limit under
consideration. We checked Eq.(\ref{e19}) additionally by a direct expansion
of the both sides in $J$ up to terms $J^2$. Eq. (\ref{e19}) was written in
Ref. \cite{mk} at $u\left({\bf r}\right) =0$ and $\hat{a}=0$ but its
accuracy remained unclear.

The form of the partition function $Z_{2}[J]$, Eqs. (\ref{sigmaaverage}, \ref
{integ}) allows us to average it immediately over both the long range
potential $u\left( {\bf r}\right) $ and the supermatrix $M\left( {\bf r}%
\right) $. Assuming that fluctuations of the random potential $u\left( {\bf r%
}\right) $ are gaussian with the correlation 
\begin{equation}
\langle u\left( {\bf r}\right) u\left( {\bf r}^{\prime }\right) \rangle
=W\left( {\bf r-r}^{\prime }\right) \text{,}  \label{e20}
\end{equation}
where $W\left( {\bf r}\right) $ is a function decaying at distances $b$ much
exceeding the wavelength $\lambda _{F}$, $b\gg \lambda _{F}$, we average the
partition function $Z_{2}\left( J\right) $ over $u\left( {\bf r}\right) $
and $M\left( {\bf r}\right) $ with the help of Eqs. (\ref{e12}), (\ref{e20})
and find for the final averaged partition function $Z\left( \hat{a}\right) $ 
\begin{equation}
Z\left( \hat{a}\right) =\int_{Q_{{\bf n}}^{2}\left( {\bf r}\right) =1}\exp
\left( -{\cal F}\left[ Q_{{\bf n}}\left( {\bf r}\right) \right] \right) DQ_{%
{\bf n}}\left( {\bf r}\right)  \label{e21}
\end{equation}
The ``free energy'' functional ${\cal F}$ takes the form 
\begin{equation}
{\cal F}\left[ Q_{{\bf n}}\left( {\bf r}\right) \right] ={\cal F}_{kin}+%
{\cal F}_{imp}+{\cal F}_{imp}^{\left( s\right) },  \label{e22}
\end{equation}
\begin{eqnarray*}
{\cal F}_{kin}[Q_{{\bf n}}({\bf r})] &=&\frac{\pi \nu }{4}Str\int d{\bf r}d%
{\bf n[}2v_{F}\Lambda {\bar{T}}_{{\bf n}}({\bf r}){\bf n}{\nabla }T_{{\bf n}%
}({\bf r}) \\
&&+2i(\frac{\omega +i\delta }{2}\Lambda -\hat{a})Q_{{\bf n}}({\bf r})]
\end{eqnarray*}

\begin{eqnarray*}
{\cal F}_{imp}\left[ Q_{{\bf n}}\right] &=&-\frac{1}{8}\left( \frac{\pi \nu 
}{p_{F}}\right) ^{2}\int d{\bf r}d{\bf n}d{\bf r}{^{\prime }}d{\bf n}{%
^{\prime }}{\bf \nabla }_{{\bf r}}^{i}{\bf \nabla }{}_{{\bf r}^{\prime
}}^{j}W({\bf r}-{\bf r}{^{\prime }}) \\
&&\times Str[\Lambda {\bar{T}}_{{\bf n}}\left( {\bf r}\right) \nabla _{{\bf n%
}}^{i}T_{{\bf n}}({\bf r})]Str[\Lambda {\bar{T}}_{{\bf n}{^{\prime }}}({\bf r%
}{^{\prime }})\nabla _{{\bf n}{^{\prime }}}^{j}T_{{\bf n}{^{\prime }}}({\bf r%
}{^{\prime }})]
\end{eqnarray*}

\[
{\cal F}_{imp}^{\left( s\right) }\left[ Q_{{\bf n}}\left( {\bf r}\right) %
\right] =-\frac{\pi \nu }{8\tau _{s}}Str\left( \int Q_{{\bf n}}\left( {\bf r}%
\right) d{\bf n}\right) ^{2} 
\]
and $Q_{{\bf n}}\left( {\bf r}\right) =T_{{\bf n}}\left( {\bf r}\right)
\Lambda {\bar{T}}_{{\bf n}}\left( {\bf r}\right) $.

As we see from Eqs. (\ref{e21}-\ref{e22}), the free energy functional ${\cal %
F}\left[ Q_{{\bf n}}\left( {\bf r}\right) \right] $ consists of three parts.

The first part ${\cal F}_{kin}\left[ Q_{{\bf n}}\left( {\bf r}\right) \right]
$ describes the kinetic modes in the absence of any impurities. Correlation
functions of interest can be obtained differentiating $Z(\hat{a})$ in $\hat{a%
}$ as written in Eqs. (\ref{e7}, \ref{e9}). The first term in ${\cal F}_{kin}%
\left[ Q_{{\bf n}}\left( {\bf r}\right) \right] $ can be written also in the
form of the Wess-Zumino-Novikov-Witten integral with an additional variable
of integration \cite{mk}. The second part ${\cal F}\left[ Q_{{\bf n}}\left( 
{\bf r}\right) \right] $ is responsible for scattering on the long range
potential, whereas the third term is due to scattering on the short range
impurities.

Eqs. (\ref{e21}-\ref{e22}) are valid even in the absence of any impurities
as long as one may use the quasiclassical approximation. This description
fails near boundaries of the sample because ``turning points'' where the
quasiclassical description fails are inevitable in those regions.
Apparently, this can lead to an additional term in the free energy
functional analogous to the regularizer introduced in Ref. \cite{al}. In all
other parts of the sample, Eqs. (\ref{e21}-\ref{e22}) are valid at all
distances exceeding the wavelength $\lambda _{F}$.

It is important to emphasize that the problem of the ``mode locking'' \cite
{asaa,zirn,aos} does not exist in the present approach. As ${\bf n}^{2}=1$,
there are no fluctuations transverse to the constant energy shell and no
additional averaging is necessary. Moreover, there are no fluctuations of
the eigenvalues of the supermatrix $Q_{{\bf n}}\left( {\bf r}\right) $
because by construction $Q_{{\bf n}}^{2}=1$. The limit of the vanishing
short range potential $\tau _{s}\rightarrow \infty $ can be taken if some
long range potential is present. The purely ballistic case without any
randomness is more sophisticated because one needs a regularizer. However,
this is a more delicate effect than the problem of the mode locking.

The term ${\cal F}_{imp}\left[ Q_{{\bf n}}\left( {\bf r}\right) \right] $,
Eq. (\ref{e22}), does not have a form of a collision integral of the
Boltzmann equation. It differs from what one obtains using the saddle-point
approximation and expanding in gradients\cite{aos,te1}. If one expands the
functional ${\cal F}\left[ Q_{{\bf n}}\left( {\bf r}\right) \right] $ in
small deviations $Q$ from $\Lambda $ using e.g. a parametrization like 
\begin{equation}
T=iP+\left( 1-P^{2}\right) ^{1/2}\text{, \ \ \ }P\Lambda +\Lambda P=0
\label{e23}
\end{equation}
the first non-vanishing contribution is of the order of $P^{4}$, which means
that ${\cal F}_{imp}\left[ Q_{{\bf n}}\left( {\bf r}\right) \right] $ does
not contribute to the bare propagator at all.

At the same time, it was demonstrated \cite{te1} that the ballistic $\sigma $%
-model with a term in the form of the collision integral could be obtained
at large distances exceeding a single particle mean free path $l$. In the
next chapter we will clarify this question by demonstrating that the
standard form of the collision integral can really be obtained at large
distances as a result of a course-graining procedure.

\section{Reduced ballistic $\protect\sigma $-model}

The free energy functional ${\cal F}\left[ Q_{{\bf n}}\left( {\bf r}\right) %
\right] $, Eqs. (\ref{e21}-\ref{e22}) is most general and applicable at all
distances exceeding the wavelength $\lambda _{F}$. We neglect now the part $%
{\cal F}_{s}\left[ Q_{{\bf n}}\left( {\bf r}\right) \right] $ originating
from the short range impurities and concentrate on studying properties of
the long range scattering.

At very large distances exceeding the mean free path $l_{tr}$ (we evaluate
this length later, see Eq.(\ref{a25})) the $\sigma $-model must acquire the
standard diffusive form\cite{efetov}. However, if the long range potential $%
u\left( {\bf r}\right) $ is weak such that $l_{tr}\gg b$, one more
intermediate scale is important for describing the behavior of the system,
namely, the one related to the Lapunov exponent $\lambda _{L}=\tau _{L}^{-1}$%
. The corresponding time $\tau _{L}$ is a time required for two particles
moving initially parallel to each other to increase the distance between
them by a factor of order unity (This definition gives, of course, the order
of magnitude of $\tau _{L}$ only). The importance of this time was first
pointed out in Ref. \cite{al}, where a weak localization correction was
calculated for scattering on a long range potential. This time is also
relevant for correlations of wave functions discussed recently\cite{gm} for
a model of weak scatterers where it was demonstrated that only at distances $%
R$ exceeding the $l_{L}=v_{F}\tau _{L}$ the notion of separate diffusons
makes a sense.

The length $l_{L}$ was estimated for the model of long range weak scatterers
corresponding to the case considered here as 
\begin{equation}
l_{L}\sim l_{tr}\left( b/l_{tr}\right) ^{2/3}  \label{a1}
\end{equation}
which shows that $l_{L}$ is between $b$ and $l_{tr}$, $b\ll l_{L}\ll l_{tr}$.

At distances $R$ $\ $smaller than $l_{L}$, two particles that started their
motion along parallel trajectories still move parallel to each other and,
following Ref. \cite{al}, we call the region inside $l_{L}$ Lapunov region.

In this chapter, we want to show that the ballistic non-linear $\sigma $%
-model, Eqs. (\ref{e21}, \ref{e22}), acquires a more familiar form \cite
{aos,te1} provided an integration over $Q_{{\bf n}}\left( {\bf r}\right) $
within the Lapunov region is performed. This integration is some kind of the
course-graining procedure used very often in statistical physics. As a
result of such an integration, we obtain another field theory that can be
called a reduced ballistic $\sigma $-model.

Proceeding in the standard way we separate fluctuations of the matrix $Q_{%
{\bf n}}({\bf r})$ into fast and slow parts. By fast variations of the
supermatrix $Q_{{\bf n}}\left( {\bf r}\right) $ we mean fluctuations
changing fast within the Lapunov region and the rest is classified as slow
ones. This procedure is very similar to the renormalization group (RG)
scheme applied in $2D$ in the diffusive region (see, e.g. \cite{efetov}).
Carrying out the course-graining procedure when deriving the RG equations is
simplified by the fact that all arising integrals are logarithmic and one
needs to know only the order of magnitude of the cutoffs at each step of the
shell integration. In the course-graining procedure used here, resulting
integrals are not logarithmic. At first glance, this would make the entire
procedure rather tricky because a renormalized free energy functional would
depend on the length $l_{L}$ estimated by the order of magnitude only.

Fortunately, there exists a method of integration over the fast modes
resulting in a renormalized free energy functional $F[\tilde{Q}_{{\bf n}}]$
Eq.(\ref{a18}) that does not contain $l_{L}$ as a parameter. This length
will be implied for $F[\tilde{Q}_{{\bf n}}]$ as an ulraviolet cutoff only.

The separation into slow and fast fluctuating parts is performed as follows

\begin{equation}
T_{{\bf n}}({\bf r})=\tilde{T}_{{\bf n}}({\bf r})V_{{\bf n}}({\bf r})
\label{separation}
\end{equation}
where both $\tilde{T}_{{\bf n}}({\bf r})$ are $V_{{\bf n}}({\bf r})$ are
unitary supermatrices. The supermatrix $\tilde{T}_{{\bf n}}({\bf r})$ is
supposed to describe slow modes, whereas $V_{{\bf n}}({\bf r})$ is
responsible for fast ones. Substitution of Eq.(\ref{separation}) into Eq.(%
\ref{e22}) results in a new action in which the two fields interact in
complicated way. We write this action as follows 
\begin{eqnarray}
{\cal F}_{kin}[T] &=&{\cal F}_{kin}\left[ V\right] +{\cal F}_{kin}^{\prime }[%
\tilde{T},V]  \label{a2} \\
{\cal F}_{imp}\left[ T\right] &=&{\cal F}_{imp}\left[ V\right] +{\cal F}%
_{imp}^{\prime }[\tilde{T},V]  \nonumber
\end{eqnarray}
where 
\begin{eqnarray}
{\cal F}_{kin}[V] &=&\frac{\pi \nu }{4}Str\int d{\bf r}d{\bf n[}%
2v_{F}\Lambda \bar{V}_{{\bf n}}({\bf r}){\bf n\nabla }V_{{\bf n}}({\bf r}) 
\nonumber \\
&&+(i\omega -2i\hat{a}\Lambda -2\lambda _{L})\Lambda Q_{{\bf n}}^{\left(
0\right) }\left( {\bf r}\right) ]  \label{a3}
\end{eqnarray}
\begin{eqnarray}
{\cal F}_{kin}^{\prime }[\tilde{T},V] &=&\frac{\pi \nu }{4}Str\int d{\bf r}d%
{\bf n[}2v_{F}Q_{{\bf n}}^{\left( 0\right) }\bar{\tilde{T}}_{{\bf n}}\left( 
{\bf r}\right) {\bf n\nabla }\tilde{T}_{{\bf n}}\left( {\bf r}\right) 
\nonumber \\
&&+i\big(\bar{\tilde{T}}_{{\bf n}}\left( {\bf r}\right) (\omega +i\delta -2%
\hat{a}\Lambda )\Lambda \tilde{T}_{{\bf n}}\left( {\bf r}\right)  \nonumber
\\
&&-(\omega +i\delta )\Lambda +2\hat{a}\big)Q_{{\bf n}}^{\left( 0\right)
}\left( {\bf r}\right) ]  \label{a4}
\end{eqnarray}
\[
{\cal F}_{imp}\left[ V\right] =-\frac{1}{8}\left( \frac{\pi \nu }{p_{F}}%
\right) ^{2}\int d{\bf r}d{\bf n}d{\bf r}{^{\prime }}d{\bf n}{^{\prime }}%
\nabla _{{\bf r}}^{i}\nabla _{{\bf r}^{\prime }}^{j}W({\bf r}-{\bf r}{%
^{\prime }}) 
\]
\begin{equation}
\times Str[\Lambda \bar{{V}}_{{\bf n}}({\bf r})\nabla _{{\bf n}}^{i}V_{{\bf n%
}}({\bf r})]Str[\Lambda \bar{{V}}_{{\bf n}{^{\prime }}}({\bf r}{^{\prime }}%
)\nabla _{{\bf n}{^{\prime }}}^{j}V_{{\bf n}{^{\prime }}}({\bf r}{^{\prime }}%
)]  \label{a5}
\end{equation}

\[
{\cal F}_{imp}^{\prime }[\tilde{T},V]=-\frac{1}{8}\left( \frac{\pi \nu }{%
p_{F}}\right) ^{2}\int d{\bf r}d{\bf r}^{\prime }d{\bf n}d{\bf n}^{\prime
}\nabla _{{\bf r}}^{i}\nabla _{{\bf r}^{\prime }}^{j}W\left( {\bf r-r}%
^{\prime }\right) 
\]
\[
\times \lbrack Str(Q_{{\bf n}}^{\left( 0\right) }\left( {\bf r}\right) \Phi
_{{\bf n}}^{i}\left( {\bf r}\right) )Str(Q_{{\bf n}^{\prime }}^{\left(
0\right) }\left( {\bf r}^{\prime }\right) \Phi _{{\bf n}^{\prime
}}^{j}\left( {\bf r}^{\prime }\right) ) 
\]
\begin{equation}
+2Str(\Lambda \bar{V}_{{\bf n}}\left( {\bf r}\right) \nabla _{{\bf n}}^{i}V_{%
{\bf n}})Str(Q_{{\bf n}^{\prime }}^{\left( 0\right) }\left( {\bf r}\right)
\Phi _{{\bf n}^{\prime }}^{j}\left( {\bf r}^{\prime }\right) )]  \label{a6}
\end{equation}

In Eqs. (\ref{a3}-\ref{a6}), 
\begin{equation}
Q_{{\bf n}}^{\left( 0\right) }=V_{{\bf n}}\left( {\bf r}\right) \Lambda \bar{%
V}_{{\bf n}}\left( {\bf r}\right) \text{, \ \ \ \ \ \ \ }\Phi _{{\bf n}%
}^{i}\left( {\bf r}\right) =\bar{\tilde{T}}_{{\bf n}}\left( {\bf r}\right)
\nabla _{{\bf n}}^{i}\tilde{T}_{{\bf n}}\left( {\bf r}\right) .  \label{a6a}
\end{equation}

The separation into the fast and slow parts, Eq. (\ref{separation}), is not
simple for the problem involved. This is because, for a given ${\bf n}$, a
characteristic dependence of $Q_{{\bf n}}\left( {\bf r}\right) $ on the
coordinate ${\bf r}$ is extremely anisotropic and one cannot introduce an
isotropic momentum shell for integration over the fast modes. Besides, we
should not violate the rotational invariance when integrating over the fast
modes. This goal is achieved by writing the term with the Lapunov exponent $%
\lambda _{L}$ in ${\cal F}_{kin}[V]$, Eq. (\ref{a3}). This is similar to an
invariant integration over a momentum shell for a diffusive $\sigma $-model.

Introducing a notation 
\[
\langle \dots \rangle _{0}\equiv \int \left( \dots \right) \exp \left( -%
{\cal F}_{kin}[V]\right) DQ_{{\bf n}}^{\left( 0\right) } 
\]
we write the reduced free energy $F$ as 
\begin{eqnarray}
&&F=-\ln \langle \exp (-{\cal F}_{kin}^{\prime }[\tilde{T},V]-  \nonumber \\
&&-{\cal F}_{imp}\left[ V\right] -{\cal F}_{imp}^{\prime }[\tilde{T}%
,V])\rangle _{0}  \label{a7}
\end{eqnarray}
(due to the supersymmetry $\langle 1\rangle _{0}=1$).

We calculate the average in Eq.(\ref{a7}) by expansion of the exponential
and computing averages of all terms obtained in this way. In the first
order, using the relation 
\begin{equation}
\langle Q_{{\bf n}}^{\left( 0\right) }\left( {\bf r}\right) \rangle
_{0}=\Lambda  \label{a7a}
\end{equation}
we obtain 
\begin{equation}
\langle {\cal F}_{kin}^{\prime }[\tilde{T},V]\rangle _{0}={\cal F}_{kin}[%
\tilde{Q}_{{\bf n}}\left( {\bf r}\right) ]  \label{a8}
\end{equation}
with ${\cal F}_{kin}[\tilde{Q}_{{\bf n}}\left( {\bf r}\right) ]$ from Eq. (%
\ref{e22}), 
\begin{equation}
\langle {\cal F}_{imp}\left[ V\right] \rangle _{0}=0,  \label{a9}
\end{equation}
which is due to the supersymmetry, as well as the second term in ${\cal F}%
_{imp}^{\prime }[\tilde{T},V]$, Eq.(\ref{a6}),

\[
\langle{\cal F}_{imp}^{\prime }[\tilde{T},V]\rangle_{0}= -\frac{1}{8}\left( 
\frac{\pi\nu}{p_{F}}\right) ^{2} \int d{\bf r}d{\bf r}^{\prime }d{\bf n}d%
{\bf n}^{\prime }\nabla_{{\bf r}}^{i}\nabla_{{\bf r}^{\prime }}^{j}W\left( 
{\bf r-r}^{\prime }\right) 
\]
\begin{equation}
\times\langle Str(Q_{{\bf n}}^{\left( 0\right) }\left( {\bf r}\right) \Phi _{%
{\bf n}}^{i}\left( {\bf r}\right) ) Str(Q_{{\bf n}^{\prime }}^{\left(
0\right)}\left( {\bf r}^{\prime }\right) \Phi _{{\bf n}^{\prime }}^{j}\left( 
{\bf r}^{\prime }\right) )\rangle_{0}  \label{a10}
\end{equation}

If we replaced the average of the product of $Q_{{\bf n}}^{\left( 0\right) }$
in Eq. (\ref{a10}) by the product of the averages we would simply come back
using Eq. (\ref{a7a}) to ${\cal F}_{imp}[\tilde{Q}_{{\bf n}}\left( {\bf r}%
\right) ]$ in Eq. (\ref{e22}). However, now the supermatrices $\tilde{Q}_{%
{\bf n}}\left( {\bf r}\right) $ vary at much longer distances than the
radius $b$ of the correlation function $W\left( {\bf r-r}^{\prime }\right) $%
. As the integrand in Eq. (\ref{a10}) contains derivatives ${\bf \nabla }%
_{r} $ of $W\left( {\bf r-r}^{\prime }\right) $, the integral is small and
can be neglected.

The main contribution comes from the irreducible part of the correlation
function which we denote as 
\begin{equation}
\langle \langle \left( Q_{{\bf n}}^{\left( 0\right) }\left( {\bf r}\right)
\right) ^{\alpha \beta }\left( Q_{{\bf n}^{\prime }}^{\left( 0\right)
}\left( {\bf r}^{\prime }\right) \right) ^{\gamma \mu }\rangle \rangle _{0}
\label{a11}
\end{equation}
(at the moment we do not distinguish between the advanced/retarded blocks
and the others and write all indices standing for matrix elements as
superscripts). We calculate this average in Chapter V in a general case
including the impurity potential (see Eq. \ref{eq.11}). Now we neglect the
impurity potential and make the replacement $\delta \rightarrow \lambda _{L}$
when determining the function ${\cal G}_{{\bf nn}{^{\prime }}}({\bf \rho })$
from Eq.(\ref{eq.10}). Then, we substitute Eq.(\ref{eq.11}) into Eq.(\ref
{a10}) and transform the result of the substitution by introducing new
integration variables ${\bf R}=\frac{{\bf r}+{\bf r}{^{\prime }}}{2}$, ${\bf %
\rho }={\bf r}-{\bf r}{^{\prime }}$. Then, we note that the integrand
contains the function ${\cal G}_{{\bf nn}{^{\prime }}}({\bf \rho })$ that
changes on the distance $l_{L}$ (the limit $\omega \ll \lambda _{L}$ is
implied), while the supermatrix $\Phi _{{\bf n}}^{i}\left( {\bf r}\right) $
changes slower. This allows us to rewrite Eq. (\ref{a10}) as follows

\begin{eqnarray}
&&\langle {\cal F}_{imp}^{\prime }[\tilde{T},V]\rangle _{0}=-\frac{\pi \nu }{%
2p_{F}^{2}}Str\int d{\bf r}d{\bf n}d{\bf n}{^{\prime }}\bigg[\big(\Phi _{%
{\bf n}}^{i}({\bf r})\big)^{\perp }  \nonumber \\
&&\times \big(\Phi _{{\bf n}^{\prime }}^{j}({\bf r})\big)^{\perp }\int d{\bf %
\rho }\nabla ^{i}\nabla ^{j}W({\bf \rho })\Lambda {\cal G}_{{\bf nn}{%
^{\prime }}}({\bf \rho })\biggr]  \label{eq.13}
\end{eqnarray}
The notation $(\dots )^{\perp }$ means here the part of the supermatrices
anticommuting with $\Lambda $. Eq. (\ref{eq.13}) can be reduced to a more
simple form. In the Fourier transformed representation, the integral $I_{%
{\bf nn}^{\prime }}$ over ${\bf \rho }$ in Eq. (\ref{eq.13}) takes the form 
\begin{equation}
I_{{\bf nn}{^{\prime }}}=-\int \frac{d{\bf q}}{(2\pi )^{d}}{\bf q}^{i}{\bf q}%
^{j}W({\bf q}){\cal G}_{{\bf nn}{^{\prime }}}(-{\bf q})  \label{eq.15}
\end{equation}
Solving Eq.(\ref{eq.10}) we find for the Green function in the limit $\omega
\ll \lambda _{L}$

\begin{equation}
{\cal G}_{{\bf nn}{^{\prime }}}(-{\bf q})=-\frac{\delta _{{\bf nn}{^{\prime }%
}}}{iv_{F}{\bf nq}+\lambda _{L}\Lambda }  \label{eq.16}
\end{equation}
The function $W\left( {\bf q}\right) $ decays at momenta much smaller than $%
p_{F}$ and only such momenta give the main contribution in Eq. (\ref{eq.15}%
). Therefore, we can approximately write the function ${\cal G}_{{\bf n}{%
{\bf n}^{\prime }}}(-{\bf q})$ in the integral, Eq.(\ref{eq.15}), as

\begin{equation}
I_{{\bf nn}{^{\prime }}}=i\delta _{{\bf nn}{^{\prime }}}\int \frac{d{\bf q}}{%
(2\pi )^{d}}q^{i}q^{j}W({\bf q})\left( \frac{({\bf p}-{\bf q})^{2}}{2m}%
-\varepsilon _{F}+i\lambda _{L}\Lambda \right) ^{-1}  \label{eq.17}
\end{equation}
where ${\bf p}=p_{F}{\bf n}$.

Using the fact that 
\[
\big(\Phi _{{\bf n}}^{i}({\bf r})\big)^{\perp} \big(\Phi_{{\bf n}^{\prime
}}^{j}({\bf r})\big)^{\perp}=-\frac{1}{4}\nabla _{{\bf n}}^{i}\tilde{Q}%
\left( {\bf r}\right) \nabla _{{\bf n}}^{j}\tilde{Q}\left( {\bf r}\right) 
\]
we reduce Eq. (\ref{eq.13}) to the form

\begin{eqnarray}
&&\left\langle {\cal F}_{imp}^{\prime }[\tilde{T},V]\right\rangle _{0}=%
\frac{i\pi \nu }{8p_{F}^{2}}Str\int d{\bf n}d{\bf r}\nabla _{{\bf n}%
}^{i}Q\left( {\bf r}\right) \nabla _{{\bf n}}^{j}Q\left( {\bf r}\right) 
\label{eq.18} \\
&&\times \int \frac{d{\bf q}}{\left( 2\pi \right) ^{d}}{\bf q}^{i}{\bf q}%
^{j}W({\bf q})\left( \frac{({\bf p}-{\bf q})^{2}}{2m}-\varepsilon
_{F}+i\lambda _{L}\Lambda \right) ^{-1}  \nonumber
\end{eqnarray}
Changing the variables of the integration in the integral Eq.(\ref{eq.18})
to ${\bf p}{^{\prime }}={\bf p}-{\bf q}$ and integrating separately over $%
\xi =p^{\prime 2}/2m-\varepsilon _{F}$ and ${\bf n}{^{\prime }}={\bf p}{%
^{\prime }/}p^{\prime }$ we obtain finally

\begin{eqnarray}
\left\langle {\cal F}_{imp}^{\prime }[\tilde{T},V]\right\rangle _{0} &=&%
\frac{\left( \pi \nu \right) ^{2}}{8}\int d{\bf r}\int d{\bf n}d{\bf n}{%
^{\prime }}({\bf n}-{\bf n}^{\prime })^{i}({\bf n}-{\bf n}^{\prime })^{j} 
\nonumber \\
&&\times W_{{\bf nn}{^{\prime }}}Str[\nabla _{{\bf n}}^{i}\tilde{Q}_{{\bf n}%
}({\bf r})\nabla _{{\bf n}}^{j}\tilde{Q}_{{\bf n}}({\bf r})]  \label{eq.20}
\end{eqnarray}
where $W_{{\bf nn}^{\prime }}=W\left( p_{F}\left( {\bf n-n}^{\prime }\right)
\right) $. This result corresponds to the collision integral in the
Boltzmann kinetic equation in the limit of small angles of the scattering.

Proceeding further we can calculate, in principle, not only $\langle {\cal F}%
_{imp}^{\prime }[\tilde{T},V]\rangle _{0}$ but also its higher cummulants
like 
\[
\left\langle \left\langle {\cal F}_{imp}^{\prime }[\tilde{T},V]{\cal F}%
_{imp}^{\prime }[\tilde{T},V]\right\rangle \right\rangle _{0}
\]
To estimate these cummulants we should consider an average of the type 
\begin{eqnarray}
&&\langle \langle \nabla _{{\bf n}}\nabla _{{\bf n}^{\prime }}Q_{{\bf n}%
}^{\left( 0\right) }\left( {\bf r}\right) Q_{{\bf n}^{\prime }}^{\left(
0\right) }\left( {\bf r}^{\prime }\right)   \nonumber \\
&&\times \nabla _{{\bf n}_{1}}\nabla _{{\bf n}_{1}^{\prime }}Q_{{\bf n}%
_{1}}^{\left( 0\right) }\left( {\bf r}_{1}\right) Q_{{\bf n}_{1}^{\prime
}}^{\left( 0\right) }\left( {\bf r}_{1}^{\prime }\right) \rangle \rangle _{0}
\label{a17}
\end{eqnarray}
where $|{\bf r-r}^{\prime }|\sim |{\bf r}_{1}{\bf -r}_{1}^{\prime }|\lesssim
b$ and we do not write explicitly indices.

The correlation function for the product of four$\ Q^{(0)}$, Eq. (\ref{a17}%
), resembles a correlation function calculated in Ref. \cite{gm} where it
was demonstrated that, being a complicated function of small distances, the
reducible correlation function decoupled into $2$ diffusons as soon as the
distance between the points ${\bf r,r}^{\prime }$ and ${\bf r}_{1},{\bf r}%
_{1}^{\prime }$ exceeded the Lapunov length $l_{L}$. This means that the
irreducible correlation function of the type as in Eq. (\ref{a17}) decays at
distances of the order of the length $l_{L}$. As the second cummulant
contains a product of four slow functions $\Phi _{{\bf n}}\left( {\bf r}%
\right) $, Eq. (\ref{a6a}), we conclude that the cummulant expansion is
effectively a series in $l_{L}\nabla _{{\bf r}}$. Therefore, beyond the
Lapunov region we may keep the average ${\cal F}_{imp}^{\prime }[\tilde{T},V]
$ only. Using Eqs. (\ref{a7}-\ref{a9}, \ref{eq.20}) we write the reduced
ballistic $\sigma $-model applicable beyond the Lapunov region in the form 
\begin{eqnarray}
F[Q_{{\bf n}}\left( {\bf r}\right) ] &=&\frac{\pi \nu }{4}Str\int d{\bf r}%
\bigg[\int (2v_{F}\Lambda \overline{T}_{{\bf n}}\left( {\bf r}\right) \nabla
_{{\bf r}}T_{{\bf n}}\left( {\bf r}\right)   \nonumber \\
&&+i(\omega +i\delta -2\hat{a}\Lambda )\Lambda Q_{{\bf n}}\left( {\bf r}%
\right) )d{\bf n}  \nonumber \\
&&+\frac{\pi \nu }{2}\int W_{{\bf nn}^{\prime }}\left( {\bf n}-{\bf n}%
^{\prime }\right) ^{i}\left( {\bf n}-{\bf n}^{\prime }\right) ^{j}  \nonumber
\\
&&\times \nabla _{{\bf n}}^{i}Q_{{\bf n}}({\bf r})\nabla _{{\bf n}}^{j}Q_{%
{\bf n}}({\bf r})d{\bf n}d{\bf n}^{\prime }\bigg]  \label{a18}
\end{eqnarray}
The reduced ballistic $\sigma $-model, Eq. (\ref{a18}), has the same form as
the corresponding $\sigma $-model derived for a long range potential \cite
{te1} provided the limit of small angle scattering is considered. The
non-linear $\sigma $-model was obtained in Ref. \cite{te1} under the
assumption that all distances exceeded the single particle mean free path $l$
using, as the first step, the saddle point approximation. However, the
saddle point approximation is not good for a long range potential and,
hence, the length $l$ cannot be a good quantity. Now we see that one should
use the length $l_{L}$. The reduced $\sigma $-model is applicable at
distances larger than $l_{L}$ and this agrees with the suggestion of Ref. 
\cite{gm}. It is remarkable, that the reduced $\sigma $-model does not
contain explicitly the length $l_{L}$ as a parameter. It can enter as an
ultraviolet cutoff only. Therefore, it is sufficient to know $l_{L}$ by
order of magnitude.

One can further simplify the reduced ballistic $\sigma $-model, Eq. (\ref
{a18}), at distances exceeding the transport mean free path $l_{tr}\gg l_{L}$%
. This route is well developed \cite{mk,asaa,te1}. Separating again fast
modes (strongly varying at distances smaller than $l_{tr}$) from slow ones
with the zero angular harmonics (fluctuating at distances exceeding $l_{tr}$%
) we write the supermatrix $T_{{\bf n}}\left( {\bf r}\right) $ in Eq. (\ref
{a18}) as 
\begin{equation}
T_{{\bf n}}\left( {\bf r}\right) =U\left( {\bf r}\right) T_{{\bf n}}^{\left(
0\right) }\left( {\bf r}\right)  \label{a19}
\end{equation}
Then, the free energy functional $F[Q_{{\bf n}}\left( {\bf r}\right) ]$, Eq.
(\ref{a18}), takes the form 
\begin{equation}
F[U,T^{\left( 0\right) }]=F_{0}[T^{\left( 0\right) }]{\bf +}F^{\prime
}[U,T^{\left( 0\right) }],  \label{a20}
\end{equation}
\begin{eqnarray}
F_{0}[T^{\left( 0\right) }] &=&\frac{\pi \nu }{4}Str\int [2v_{F}\overline{T}%
_{{\bf n}}^{\left( 0\right) }\left( {\bf r}\right) {\bf n}\nabla _{{\bf r}%
}T_{{\bf n}}^{\left( 0\right) }\left( {\bf r}\right)  \nonumber \\
&&+i(\omega +i\delta )\Lambda Q_{{\bf n}}^{\left( 0\right) }\left( {\bf r}\right) ]d%
{\bf r}  \label{a21a}
\end{eqnarray}

\begin{equation}
F^{\prime }\left[ U,T^{\left( 0\right) }\right] =\frac{\pi \nu }{4}Str\int d%
{\bf r}[2v_{F}Q_{{\bf n}}^{\left( 0\right) }\left( {\bf r}\right) {\bf n}%
\Phi \left( {\bf r}\right)  \label{a21}
\end{equation}
\[
+i\left( \omega +i\delta \right) \left( \bar{U}\left( {\bf r}\right) \Lambda
U\left( {\bf r}\right) -\Lambda \right) Q_{{\bf n}}^{\left( 0\right) }\left( 
{\bf r}\right) + 
\]
\[
\frac{\pi \nu }{2}\int W_{{\bf nn}^{\prime }}\left( {\bf n-n}^{\prime
}\right) ^{i}\left( {\bf n-n}^{\prime }\right) ^{j}\nabla _{{\bf n}}^{i}Q_{%
{\bf n}}^{\left( 0\right) }({\bf r})\nabla _{{\bf n}}^{j}Q_{{\bf n}}^{\left(
0\right) }({\bf r})d{\bf n}d{\bf n}^{\prime }] 
\]
where $Q_{{\bf n}}^{\left( 0\right) }\left( {\bf r}\right) =T_{{\bf n}%
}^{\left( 0\right) }\left( {\bf r}\right) \Lambda \bar{T}_{{\bf n}}^{\left(
0\right) }\left( {\bf r}\right) $, $\Phi \left( {\bf r}\right) =\bar{U}%
\left( {\bf r}\right) \nabla U\left( {\bf r}\right) $.

The next step is to integrate over the fast modes, which leads to the free
energy functional $F_{dif}\left[ Q\right] $%
\begin{equation}
F_{dif}\left[ Q\right] =\langle \exp \left( -F^{\prime }\left[ U,T^{\left(
0\right) }\right] \right) \rangle _{0}  \label{a22}
\end{equation}
where 
\[
<...>_{0}=\int \left( ...\right) \exp \left( -F_{0}\left[ T^{\left( 0\right)
}\right] \right) DQ^{\left( 0\right) } 
\]
Integration over $Q_{{\bf n}}^{\left( 0\right) \text{ }}\left( {\bf r}%
\right) $ can be performed using, e.g. the parametrization 
\begin{equation}
Q_{{\bf n}}^{\left( 0\right) }\left( {\bf r}\right) =\Lambda \left( 1+iP_{%
{\bf n}}\left( {\bf r}\right) \right) \left( 1-iP_{{\bf n}}\left( {\bf r}%
\right) \right) ^{-1}  \label{a23}
\end{equation}
with $\bar{P}_{{\bf n}}=-P_{{\bf n}}$, $P_{{\bf n}}\Lambda +\Lambda P_{{\bf n%
}}=0$.

Then, one can expand $P_{{\bf n}}\left( {\bf r}\right) $ (as a function of $%
{\bf n}$) in angular harmonics and expand in $P_{{\bf n}}$ in Eqs. (\ref
{a21a}). In the first term in Eq. (\ref{a21}), it is sufficient to keep only
the linear term in $P_{{\bf n}}$, while in the third term we should expand up to
quadratic terms. Due to the presence of the linear term integration over the
first harmonics is most important. Performing this gaussian integration we
come to the diffusive $\sigma $-model 
\begin{equation}
F_{dif}\left[ Q\right] =\frac{\pi \nu }{8}Str\int [D\left( \nabla Q\right)
^{2}+2i\left( \omega +i\delta \right) \Lambda Q]d{\bf r}  \label{a24}
\end{equation}
where $Q\left( {\bf r}\right) =U\left( {\bf r}\right) \Lambda \bar{U}\left( 
{\bf r}\right) $ and $D=v_{F}^{2}\tau _{tr}/d$, $d$ is the dimensionality.

The transport time $\tau _{tr}$ can be written through the function $W_{{\bf %
nn}^{\prime }}$ as 
\begin{equation}
\tau _{tr}^{-1}=2\pi \nu \int W_{{\bf nn}^{\prime }}\left( 1-{\bf nn}%
^{\prime }\right) d{\bf n}^{\prime }  \label{a25}
\end{equation}
whereas the transport mean free path $l_{tr}$ is $l_{tr}=v_{F}\tau _{tr}$.
If in addition the short range potential is present, the effective mean free
time $\tau _{eff}$ can be written as $\tau _{eff}^{-1}=\tau _{tr}^{-1}+\tau
_{s}^{-1}$.

Writing the reduced $\sigma $-model, Eq. (\ref{a18}), in a limited volume
one can come rather easily to the Wigner-Dyson level-level correlation
functions \cite{mehta}. This can be done separating fluctuations of $T_{{\bf %
n}}$ with the zero space and angle harmonics from other degrees of freedom.
We write this separation in a form similar to Eqs. (\ref{separation}, \ref
{a19}) 
\begin{equation}
T_{{\bf n}}\left( {\bf r}\right) =UT_{{\bf n}}^{\prime }\left( {\bf r}%
\right)   \label{a26}
\end{equation}
where $U$ depends neither on the vector ${\bf n}$ nor on the coordinate $%
{\bf r}$, while $T_{{\bf n}}^{\prime }\left( {\bf r}\right) $ contains only
non-zero harmonics in the both coordinates. Integration over $T_{{\bf n}%
}^{\prime }\left( {\bf r}\right) $ can be performed using a parametrization
analogous to Eq. (\ref{a23}). Excitations corresponding to the matrix $T_{%
{\bf n}}^{\prime }\left( {\bf r}\right) $ have a gap and their contribution
can be neglected in the collision region in the main approximation in the
parameter $\omega min \left( \tau _{tr},\tau _{L}\right) $. Then, we come
to the zero-dimensional version $F_{0}[Q]$ of the $\sigma $-model 
\begin{equation}
F_{0}[Q]=\frac{\pi i\left( \omega +i\delta \right) }{4\Delta }Str\left(
\Lambda Q\right)   \label{a27}
\end{equation}
where $\Delta =\left( \nu V\right) ^{-1}$ ($V$ is the volume) is the mean
level spacing. This leads directly to the Wigner-Dyson statistics \cite
{efetov}.

The above discussion describes completely the method of derivation of the $%
\sigma $-model based on quasiclassical Green functions in all regions.
However, explicit calculations with the ballistic $\sigma $-model are not
simple and calculational schemes are not necessarily the same as those used
in the diffusive limit. The most unusual is the Lapunov region. In the next
chapters we consider calculational schemes that can be useful for clean
systems.

\section{Correlation functions for a long range disorder}

The ballistic $\sigma $-model derived in the previous chapters has a form of
a functional integral over the supermatrix $Q_{{\bf n}}\left( {\bf r}\right) 
$ with the constraint $Q_{{\bf n}}^{2}\left( {\bf r}\right) =1$. Using
parametrizations of the type of Eqs. (\ref{e23}, \ref{a23}) and expanding in 
$P_{{\bf n}}$ is a potentially dangerous procedure because the rotational invariance
may be lost. From study of disordered metals \cite{efetov} we know that such
a perturbation theory works well in the diffusive region but fails in the
regime of strong localization. \ Interestingly enough, the perturbation
theory in terms of the ballistic excitations does not work also in the
Lapunov region. Using the expansion in $P_{{\bf n}}$ one comes to propagators 
\begin{equation}
\left( v_{F}{\bf nq}+\omega \Lambda \right) ^{-1}  \label{a100}
\end{equation}
and integrals over both ${\bf n}$ and vectors ${\bf q}$.

These integrals are not generally convergent and the result depends strongly
on a regularization. In Ref. \cite{asaa} the regularization was carried out
by writing an additional term in the $\sigma $-model that would arise if a
small amount of short range impurities was added. It was suggested that, at
the end of calculations, the regularizer could be put to zero. In contrast,
the authors of Ref. \cite{al} assumed that a regularizer had to remain
finite and demonstrated in a subsequent publication \cite{al1} that its
presence in the $\sigma $-model leads to an anomalous contribution. Problems
with diagrammatic expansions in the ballistic excitations were also
discussed in Ref. \cite{lerner}. The fact that one can hardly speak about
separate diffusons (kinetons) in the Lapunov region has been noticed in Ref. 
\cite{al} and emphasized in Ref. \cite{gm}.

At the same time, the integration over $Q_{{\bf n}}$ in the ballistic $%
\sigma $-model, Eq. (\ref{e21}, \ref{e22}) is well defined and the problems
arise only after using a parametrization like those in Eqs. (\ref{e23}, \ref
{a23}) and a subsequent expansion in $P_{{\bf n}}$. Therefore, we should understand
what one can do if one may not use such a parametrization. It turns out that
the most proper way of computations of correlation functions is deriving
equations for them. This is analogous to what one does in models for
turbulence \cite{falk}. Unfortunately, this method is not as general as the
perturbation theory but for some correlation functions closed equations can
be derived without difficulties.

As an example, we consider the function $Y^{00}({\bf r}_{1}{\bf ,r}%
_{2};\omega )$, Eq. (\ref{e7a}) (without averaging over the long range
potential). Using Eqs. (\ref{e9}, \ref{e19}) we reduce this correlation
function to the form 
\[
Y^{00}({\bf r}_{1}{\bf ,r}_{2};\omega )=-2(\pi \nu )^{2}\int d{\bf n_{1}}d%
{\bf n_{2}} 
\]
\begin{equation}
\times \int_{Q_{{\bf n}}^{2}=1}Q_{{\bf n_{1}}}^{84}({\bf r_{1}})Q_{{\bf n_{2}%
}}^{48}({\bf r_{2}})\exp \left( -\frac{\pi \nu }{2}\Phi _{0}[Q_{{\bf n}}(%
{\bf r})]\right) DQ_{{\bf n}}  \label{b1}
\end{equation}
with the functional $\Phi _{0}[Q_{{\bf n}}({\bf r})]$ from Eq. (\ref
{sigmaaverage}) in which we neglect the short range impurities and put $J=0$.

In order to find the average $\langle Q_{{\bf n_{1}}}^{84}({\bf r_{1}})Q_{%
{\bf n_{2}}}^{48}({\bf r_{2}})\rangle _{Q}$ (here and below the symbol $%
\langle \dots \rangle _{Q}$ is used for averaging with the functional $\Phi
_{0}[Q_{{\bf n}}({\bf r})]$) we introduce the function $g_{{\bf n}}\left( 
{\bf r};\hat{a}_{{\bf n}}\right) $ (superscripts are omitted): 
\begin{equation}
g_{{\bf n}}\left( {\bf r};\hat{a}_{{\bf n}}\right) =Z_{0}^{-1}[\hat{a}_{{\bf %
n}}]\int_{Q_{{\bf n}}^{2}=1}Q_{{\bf n}}\left( {\bf r}\right) \exp \left( -%
\frac{\pi \nu }{2}\Phi _{\hat{a}_{{\bf n}}}[Q_{{\bf n}}]\right) DQ_{{\bf n}}
\label{a12}
\end{equation}
\[
Z_{0}[\hat{a}_{{\bf n}}]=\int_{Q_{{\bf n}}^{2}=1}\exp \left( -\frac{\pi \nu 
}{2}\Phi _{\hat{a}_{{\bf n}}}[Q_{{\bf n}}]\right) DQ_{{\bf n}}
\]
\begin{equation}
\Phi _{\hat{a}_{{\bf n}}}[Q_{{\bf n}}]=\Phi _{0}[Q_{{\bf n}}]-iStr\int d{\bf %
r}d{\bf n}\hat{a}_{{\bf n}}({\bf r})Q_{{\bf n}}({\bf r})  \label{a12a}
\end{equation}
Below, the source $\hat{a}_{{\bf n}}$ is assumed to be a matrix
anticommuting with $\Lambda $. Besides, in contrast to Eq.(\ref{sigmaaverage}%
), it includes now the dependence on the direction ${\bf n}$ and is no
longer self-conjugate: $\bar{\hat{a}}_{{\bf n}}({\bf r})\neq \hat{a}_{{\bf n}%
}({\bf r})$. Then, the first order $g_{{\bf n}}^{(1)}({\bf r};\hat{a}_{{\bf n%
}})$ of the expansion of the function $g_{{\bf n}}\left( {\bf r};\hat{a}_{%
{\bf n}}\right) $ in the source gives the irreduceable correlation function 
\begin{eqnarray}
g_{{\bf n}}^{\left( 1\right) }\left( {\bf r};\hat{a}_{{\bf n}}\right)  &=&i%
\frac{\pi \nu }{2}\int d{\bf r}^{\prime }d{\bf n}^{\prime }\langle \langle
Q_{{\bf n}}\left( {\bf r}\right)   \nonumber \\
&&\times Str\left( Q_{{\bf n}^{\prime }}\left( {\bf r}^{\prime }\right) 
\hat{a}_{{\bf n}^{\prime }}\left( {\bf r}^{\prime }\right) \right) \rangle
\rangle _{Q}  \label{a14}
\end{eqnarray}
At the same time, the function $g_{{\bf n}}\left( {\bf r};\hat{a}_{{\bf n}%
}\right) $, Eq.(\ref{a12a}), is a solution for the equation (c.f. with Eq. (%
\ref{eqklassfunc})) 
\begin{eqnarray}
&&\left( v_{F}{\bf n\nabla }_{{\bf r}}-p_{F}^{-1}{\bf \nabla }_{{\bf r}}u(%
{\bf r}){\bf \partial }_{{\bf n}}\right) g_{{\bf n}}({\bf r};\hat{a}_{{\bf n}%
})+  \nonumber \\
&&\frac{i(\omega +i\delta )}{2}\left[ \Lambda ,g_{{\bf n}}({\bf r};\hat{a}_{%
{\bf n}})\right] =  \nonumber \\
&=&\frac{i}{2}\left[ \hat{a}_{{\bf n}}\left( {\bf r}\right) +\bar{\hat{a}}_{%
{\bf n}}\left( {\bf r}\right) ,g_{{\bf n}}({\bf r};\hat{a}_{{\bf n}})\right] 
\label{a15}
\end{eqnarray}
Eq.(\ref{a15}) can be solved by iterations expanding in $\hat{a}_{{\bf n}}(%
{\bf r})$. As the zero order in $\hat{a}_{{\bf n}}({\bf r})$ we may take $g_{%
{\bf n}}^{\left( 0\right) }\left( {\bf r}\right) =\Lambda $ using again the
supersymmetry of the integral Eq.(\ref{a12}) when the source is disregarded.
Then, we obtain in the first order 
\begin{eqnarray}
g_{{\bf n}}^{\left( 1\right) }\left( {\bf r};\hat{a}_{{\bf n}}\right)  &=&%
\frac{i}{4}\int d{\bf r}^{\prime }d{\bf n}^{\prime }({\cal G}_{{\bf nn}%
^{\prime }}\left( {\bf r,r}^{\prime }\right)   \nonumber \\
&&+\Lambda {\cal G}_{{\bf nn}^{\prime }}\left( {\bf r,r}^{\prime }\right)
\Lambda )\left[ \hat{a}_{{\bf n}^{\prime }}\left( {\bf r}^{\prime }\right) +%
\bar{\hat{a}}_{{\bf n}}\left( {\bf r}\right) ,\Lambda \right]   \label{a16}
\end{eqnarray}
The kernel ${\cal G}_{{\bf nn}{^{\prime }}}({\bf r},{\bf r}{^{\prime }})$ in
this expression is a Green function for the equation: 
\begin{eqnarray}
&&\left( v_{F}{\bf n\nabla }_{{\bf r}}-p_{F}^{-1}{\bf \nabla }_{{\bf r}}u(%
{\bf r}){\bf \partial }_{{\bf n}}\right) {\cal G}_{{\bf nn}{^{\prime }}}(%
{\bf r},{\bf r}{^{\prime }})+  \nonumber \\
&&+i(\omega +i\delta )\Lambda {\cal G}_{{\bf nn}{^{\prime }}}({\bf r},{\bf r}%
{^{\prime }})=\delta _{{\bf nn}{^{\prime }}}\delta ({\bf r}-{\bf r}{^{\prime
}})  \label{eq.10}
\end{eqnarray}

>From Eqs.(\ref{a14}-\ref{a16}) we find easily 
\begin{eqnarray}
&&\langle \langle \big(Q_{{\bf n}}({\bf r})\big)^{\beta \alpha }\big(Q_{{\bf %
n}^{\prime }}({\bf r}{^{\prime }})\big)^{\gamma \mu }\rangle \rangle _{Q}=%
\frac{1}{2\pi \nu }\left[ {\cal G}_{{\bf nn}{^{\prime }}}^{\beta \mu }({\bf r%
},{\bf r}{^{\prime }})\Lambda ^{\gamma \alpha }+\right.  \nonumber \\
&&\left. +\Lambda ^{\beta \nu }{\cal G}_{{\bf nn}{^{\prime }}}^{\nu \rho }(%
{\bf r},{\bf r}{^{\prime }})\Lambda ^{\rho \mu }\Lambda ^{\gamma \alpha }-%
{\cal G}_{{\bf nn}{^{\prime }}}^{\beta \nu }({\bf r},{\bf r}{^{\prime }}%
)\Lambda ^{\nu \mu }\delta ^{\gamma \alpha }-\right.  \nonumber \\
&&\left. -\Lambda ^{\beta \nu }{\cal G}_{{\bf nn}{^{\prime }}}^{\nu \mu }(%
{\bf r},{\bf r}{^{\prime }})\delta ^{\gamma \alpha }\right] K^{\gamma \gamma
}+c.c\text{.}  \label{eq.11}
\end{eqnarray}
where $c.c.$ stands for ``charge conjugated'' terms. Then, taking in Eq.(\ref
{eq.11}) different values of the superscripts one can find corresponding
integrals. For example, putting $\alpha =\gamma =4$, $\beta =\mu =8$ results
in the relation 
\begin{equation}
\langle Q_{{\bf n_{1}}}^{84}({\bf r_{1}})Q_{{\bf n_{2}}}^{48}({\bf r_{2}}%
)\rangle _{Q}=-\frac{2}{\pi \nu }{\cal D}(1;2)  \label{liouvillion}
\end{equation}
and we obtain finally 
\begin{equation}
Y^{00}({\bf r}_{1}{\bf ,r}_{2},\omega )=4\pi \nu \int d{\bf n_{1}}d{\bf n_{2}%
}{\cal D}(1;2)  \label{diffuson}
\end{equation}
The function ${\cal D}(1;2)$ used in Eqs.(\ref{liouvillion},\ref{diffuson})
is the Green function of the Liouville operator $\hat{{\cal L}}=\frac{%
\partial H}{\partial {\bf p}}\frac{\partial }{\partial {\bf r}}-\frac{%
\partial H}{\partial {\bf r}}\frac{\partial }{\partial {\bf p}}$ and
satisfies the equation 
\begin{equation}
\left( v_{F}{\bf n_{1}}{\bf \nabla _{r_{1}}}-p_{F}^{-1}{\bf \nabla _{r_{1}}}%
u({\bf r_{1}}){\bf \partial }_{{\bf n_{1}}}-i\left( \omega +i\delta \right)
\right) {\cal D}\left( 1;2\right) =\delta _{1,2}  \label{b2}
\end{equation}
We used here the notations $j\equiv ({\bf n_{j}},{\bf r_{j}})$, ${\bar{j}}%
\equiv (-{\bf n_{j}},{\bf r_{j}})$ of Ref.\cite{al}.

The quantity $Y^{00}({\bf r}_{1}{\bf ,r}_{2},\omega )$, Eq.(\ref{diffuson}),
is an extension of the non-averaged density-density correlation function to
arbitrary scales. We emphasize that the result is exact within the
quasiclassical approximation.

A similar computation can be carried out for a higher order correlation
function $Y_{2}^{00}\left( {\bf r}_{1},{\bf r}_{2},{\bf r}_{3},{\bf r}%
_{4};\omega \right) $%
\begin{eqnarray}
&&Y_{2}^{00}\left( {\bf r}_{1},{\bf r}_{2},{\bf r}_{3},{\bf r}_{4};\omega
\right)  \label{b3} \\
&=&G_{\varepsilon }^{R}\left( {\bf r}_{1},{\bf r}_{2}\right) G_{\varepsilon
-\omega }^{A}\left( {\bf r}_{2},{\bf r}_{1}\right) G_{\varepsilon
}^{R}\left( {\bf r}_{3},{\bf r}_{4}\right) G_{\varepsilon -\omega
}^{A}\left( {\bf r}_{4,}{\bf r}_{3}\right)
\end{eqnarray}
This function can be written as an integral over the supervectors $\psi
\left( {\bf r}\right) $ as

\begin{eqnarray}
&&Y_{2}^{00}\left( {\bf r}_{1},{\bf r}_{2},{\bf r}_{3},{\bf r}_{4};\omega
\right)  \label{DoubleDiffuson_psi} \\
&=&16\int \psi ^{4}({\bf r_{2}}){\bar{\psi}}^{4}({\bf r_{1}})\psi ^{8}({\bf %
r_{1}}){\bar{\psi}}^{8}({\bf r_{2}})  \nonumber \\
&&\times \psi ^{2}({\bf r_{4}}){\bar{\psi}}^{2}({\bf r_{3}})\psi ^{6}({\bf %
r_{3}}){\bar{\psi}}^{6}({\bf r_{4}})e^{-L[\psi ]}D\psi
\end{eqnarray}

Using the results of Chap. III we rewrite the integral over $\psi \left( 
{\bf r}\right) $, Eq. (\ref{DoubleDiffuson_psi}), in terms of an integral
over $Q_{{\bf n}}\left( {\bf r}\right) $ as

\begin{eqnarray}
&&Y_{2}^{00}\left( {\bf r}_{1},{\bf r}_{2},{\bf r}_{3},{\bf r}_{4};\omega
\right) =(2\pi \nu )^{2}\prod_{i=1}^{4}\int d{{\bf n}_{i}}{\cal M}(1,4;2,3)
\\
&&{\cal M}(1,4;2,3)=  \nonumber \\
&&-\left( \frac{\pi \nu }{2}\right) ^{2}\langle Q_{{\bf n_{1}}}^{84}({\bf %
r_{1}})Q_{{\bf n_{2}}}^{48}({\bf r_{2}})Q_{{\bf n_{3}}}^{62}({\bf r_{3}})Q_{%
{\bf n_{4}}}^{26}({\bf r_{4}})\rangle _{Q}  \label{DoubleDiffuson_sigma}
\end{eqnarray}
In order to calculate the correlation function ${\cal M}$, Eq. (\ref
{DoubleDiffuson_sigma}), one should expand Eqs.(\ref{a12}), (\ref{a15}) up
to the third order in the source and then compare them with each other.
Writing the integral Eq.(\ref{a12}) for $\alpha =8$, $\beta =4$ and assuming
that only the elements $\hat{a}_{{\bf n}}^{48}({\bf r})$, $\hat{a}_{{\bf n}%
}^{26}({\bf r})$, $\hat{a}_{{\bf n}}^{62}({\bf r})$ of the source are not
equal to zero we expand it in the element $\hat{a}_{{\bf n}}^{48}({\bf r})$.
Keeping only the first order in the expansion and putting $\hat{a}_{{\bf n}%
}^{48}({\bf r})=i\delta _{{\bf n_{1}}{\bf n_{2}}}\delta ({\bf r_{1}}-{\bf %
r_{2}})$ we obtain the following integral generalizing Eq.(\ref{liouvillion}%
) 
\begin{eqnarray}
&&{\cal D}_{{\bf n_{1}}{\bf n_{2}}}({\bf r_{1}},{\bf r_{2}};\hat{a}_{{\bf n}%
}^{\prime })=-\frac{\pi \nu }{2}\int_{Q_{{\bf n}}^{2}=1}Q_{{\bf n_{1}}}^{84}(%
{\bf r_{1}})Q_{{\bf n_{2}}}^{48}({\bf r_{2}})  \nonumber \\
&&\times e^{-\frac{\pi \nu }{2}\Phi _{\hat{a}_{{\bf n}}^{\prime }}[Q_{{\bf n}%
}({\bf r})]}DQ_{{\bf n}}  \label{DoubleAuxDiff}
\end{eqnarray}
The new source $\hat{a}_{{\bf n}}^{\prime }$ in Eq. (\ref{DoubleAuxDiff})
differs from the previous one $\hat{a}_{{\bf n}}$ by the substitution $\hat{a%
}_{{\bf n_{2}}}^{48}({\bf r_{2}})=0$. At the same time, making the same for
the solution of Eq.(\ref{a15}) we obtain 
\begin{eqnarray}
&&[v_{F}{\bf n_{2}}{\bf \nabla _{2}}-p_{F}^{-1}{\bf \nabla _{r_{2}}}u({\bf %
r_{2}}){\bf \partial }_{{\bf n_{2}}}  \nonumber \\
&&+i(\omega +i\delta )]{\cal D}_{{\bf n_{1}}{\bf n_{2}}}({\bf r_{1}},{\bf %
r_{2}};\hat{a}_{{\bf n}}^{\prime })  \nonumber \\
= &&\frac{1}{2}\delta _{12}[g_{{\bf n_{2}}}^{88}({\bf r_{2}};\hat{a}_{{\bf n}%
}^{\prime })-g_{{\bf n_{2}}}^{44}({\bf r_{2}};\hat{a}_{{\bf n}}^{\prime
})]Z_{2}[\hat{a}_{{\bf n}}^{\prime }]  \label{DoubleAuxEq.2}
\end{eqnarray}
In order to come now to the integral, Eq.(\ref{DoubleDiffuson_sigma}), one
should find the term of the expansion ${\cal D}_{{\bf n_{1}}{\bf n_{2}}}(%
{\bf r_{1}},{\bf r_{2}};\hat{a}_{{\bf n}}^{\prime })$ in the source $\hat{a}%
_{{\bf n}}^{\prime }$ bilinear in the elements $\hat{a}_{{\bf n}}^{26}({\bf r%
})$, $\hat{a}_{{\bf n}}^{62}({\bf r})$. As concerns the elements $g_{{\bf %
n_{2}}}^{44}({\bf r_{2}};\hat{a}_{{\bf n}}^{\prime })$, $g_{{\bf n_{2}}%
}^{88}({\bf r_{2}};\hat{a}_{{\bf n}}^{\prime })$, they satisfy Eq. (\ref{a15}%
) implying the replacement $\hat{a}_{{\bf n}}({\bf r})$ by $\hat{a}_{{\bf n}%
}^{\prime }({\bf r})$. Due to the structure of the source each of the
equations is closed and contains no terms with the source. Their solutions
are therefore $\pm 1$, respectively. Expanding further $Z_{2}[\hat{a}_{{\bf n%
}}^{\prime }]$ in the elements $\hat{a}_{{\bf n}}^{26}({\bf r})$, $\hat{a}_{%
{\bf n}}^{62}({\bf r})$ and substituting the result in Eq.(\ref
{DoubleAuxEq.2}) we obtain 
\begin{eqnarray}
&&[v_{F}{\bf n_{2}}{\bf \nabla _{2}}-p_{F}^{-1}{\bf \nabla _{r_{2}}}u({\bf %
r_{2}}){\bf \partial }_{{\bf n_{2}}}  \nonumber \\
- &&i(\omega +i\delta )]{\cal M}(1,4;{\bar{2}},3)=\delta _{1{\bar{2}}}{\cal D%
}(3;4)  \label{DoubleDiffEq1}
\end{eqnarray}
The operator in the l.h.s. of Eq.(\ref{DoubleDiffEq1}) acts on the variables
of a one of the ``diffusons''. In order to get also an equation with respect
to the variables of the other diffuson we consider the integral, Eq.(\ref
{a12}), with the source having now only the elements $\hat{a}_{{\bf n}}^{62}(%
{\bf r})$, $\hat{a}_{{\bf n}}^{48}({\bf r})$, $\hat{a}_{{\bf n}}^{84}({\bf r}%
)$. Putting there $\alpha =6$, $\beta =2$ and repeating successively all the
steps that lead us to Eqs.(\ref{DoubleAuxEq.2})-(\ref{DoubleDiffEq1}) we
come to an equation like Eq.(\ref{DoubleDiffEq1}) in which ${\bf n_{2}}$, $%
{\bf r_{2}}$ in the left side are replaced by ${\bf n_{3}}$, ${\bf r_{3}}$
and the variables $1$, ${\bar{2}}$ in the right side by $3$, $4$
respectively. Adding this equation with Eq.(\ref{DoubleDiffEq1}) we find

\begin{eqnarray}
&&(v_{F}[{\bf n_{2}}{\bf \nabla _{r_{2}}}+{\bf n_{3}}{\bf \nabla _{r_{3}}}%
]-p_{F}^{-1}[{\bf \nabla _{r_{2}}}u({\bf r_{2}}){\bf \partial}_{{\bf n_{2}}}+%
{\bf \nabla _{r_{3}}}u({\bf r_{3}}){\bf \partial}_{{\bf n_{3}}}]  \nonumber
\\
&&-2i(\omega +i\delta )){\cal M}(1,4;{\bar{2}},3)=\delta _{1{\bar{2}}}{\cal D%
}(3;4)+\delta _{34}{\cal D}(1;{\bar{2}})  \label{DoubleDiffusonEq}
\end{eqnarray}

Eq. (\ref{DoubleDiffusonEq}) agrees with the corresponding equation of Ref. 
\cite{al}. This equation is written before averaging over the long range
potential and is again exact within the quasiclassical approximation.
However, the averaging over the long range potential is not trivial.
Analyzing this equation Aleiner and Larkin \cite{al} demonstrated that the
averaged function ${\cal M}(1,4;2,3)$ decouples into $2$ diffusons (one can
connect the Green functions $G^{R,A}$ in $2$ different ways forming $2$
diffusons) at lengths exceeding the Lapunov length $l_{L}$. We call here
this region collisional. A similar analysis starting from a different
formulation was carried our in Ref. \cite{gm} where the authors came to the
same conclusion. This corresponds to the possibility of expanding in small
fluctuations of the supermatrix $Q_{{\bf n}}$ using a parametrization like
the one given by Eqs. (\ref{e23}) or (\ref{a23}).

In the Lapunov region, such an expansion is impossible and the only way to
analyze correlation functions is to write equations for them for a fixed
potential and then average the solution over the potential. Unfortunately,
we do not know how to derive exact equations for averaged correlation
functions. The form of the ballistic $\sigma $-model, Eq. (\ref{e21}, \ref
{e22}), does not allow us to derive such equations because of the
complicated form of the term ${\cal F}_{imp}$.

Apparently, using the ballistic $\sigma $-model for calculations in the
Lapunov region does not bring considerable advantages. At the same time, we
see that the $\sigma $-model is applicable also in this region, although one
cannot use the perturbation theory. Therefore, the conclusion of Ref. \cite
{gm} that a completely different theory should be constructed for the
Lapunov region is too pessimistic.

\section{Periodic orbits}

In this chapter, we consider a clean finite system (quantum billiard). In
principle, there can be holes (antidots) in the system but this can be
discussed in terms of a more complicated surface.

A standard tool for computation of the one particle density of states is the
Gutzwiller trace formula\cite{gutz} (for a review, see, e.g. \cite
{gutz,haake,gian}). This formula allows one to express the non-averaged
density of states in terms of a sum over periodic orbits. This sum is
actually divergent but there are methods to obtain reasonable results from
it. As long as energies involved are of the order of the inverse period of
short orbits one can extract a detailed information and compute also
level-level correlation functions. Very often a statistical information is
of the main interest and one calculates quantities averaged over spectrum.

The situation becomes considerably more difficult if one studies behavior at
small energies of the order of the mean level spacing $\Delta $. Even
calculation of the first non-vanishing non-oscillating terms for the
Wigner-Dyson statistics is not simple \cite{ais} and difficulties grow when
calculating next orders \cite{sieb}. On the other hand, the Wigner-Dyson
statistics can easily be obtained from the zero-dimensional $\sigma $-model
and the only question is when this $0D$ $\sigma $-model description is valid.

In Chap. IV we came to the conclusion that the $0D$ $\sigma $-model can be
used for the model of weak long range scatterers if the frequency $\omega $
is smaller than $min \left( \tau _{L}^{-1},\tau _{tr}^{-1}\right) $. For
the level-level correlation function $R\left( \omega \right) $, Eq. (\ref{e7}%
), the parameter $\omega $ is the energy difference between two levels but
what are the parameters $\tau _{tr}$ and $\tau _{L}$ or, respectively, $%
l_{tr}=v_{F}\tau _{tr}$ and $l_{L}=v_{F}\tau _{L}$?

For a clean quantum billiard the transport mean free path should be of order
of the system size $L$. At the same time, an important role should be played
by the Ehrenfest time 
\begin{equation}
t_{E}=\lambda ^{-1}\ln \left( L/\lambda _{F}\right)  \label{c1}
\end{equation}
where $\lambda $ is the Lapunov exponent for scattering on the boundaries.
The corresponding length $l_{E}=v_{F}t_{E}$ is much larger or of the order
of the system size $L$. This time determines the crossover from the
classical to the quantum regime \cite{lo,zas}. The question about the
Ehrenfest time is becoming popular in mesoscopic physics\cite{aal,been,av}.
In the present work, we are not able to obtain this time within the
approximations used. Averaging over the spectrum is equivalent to averaging
over infinite range impurities and we see from Eqs. (\ref{e21}, \ref{e22})
that only the term ${\cal F}_{kin}[Q_{{\bf n}}({\bf r})]$ is present in this
case.

At first glance, it was not necessary to average over the energy when
deriving the quasiclassical equations and the ballistic $\sigma $-model in
Chap. II. However, this depends on what limit is taken first: the infinite 
size of the systems at finite disorder or vanishing disorder in a finite 
system. In the former case, an additional averaging over the
energy is really not necessary. However, energy levels of a finite system
are quantized and we cannot directly follow the arguments of Chap. II for
the latter case. For example, the function $g_{{\bf n}}\left( {\bf r}\right) 
$ introduced in Eq. (\ref{klassaver}) is not a smooth function because one
should sum over $\xi $ instead of integrating over it. However, the
averaging over the energy improves the situation and makes possible using
the quasiclassical equations. Since we should average the partition function
with the sources, Eq. (\ref{e10}), and, hence, the Green functions, one can
simply add averaging over the energy to the summation over $\xi $ in the
definition of $g_{{\bf n}}\left( {\bf r}\right) $, Eq. (\ref{klassaver}).
Then, the function $g_{{\bf n}}\left( {\bf r}\right) $ is a smooth function
of ${\bf r}$ \ and we can repeat all the subsequent arguments leading to the 
$\sigma $-model. So, although there is no disorder in the free energy
functional ${\cal F}_{kin}[Q_{{\bf n}}({\bf r})]$, the energy averaging is
implied for the clean quantum billiards.

As we have mentioned, our quasiclassical approach should be valid everywhere
except in the vicinity of the boundaries, where the approximation fails near
the turning points. Therefore, a more accurate computation might produce an
additional term in the ballistic $\sigma $-model near the boundary. Aleiner
and Larkin \cite{al} introduced such a term modelling the quantum
diffraction by fictitious short range impurities.

We think that an additional term in the $\sigma $-model due to the quantum
diffraction is really necessary but leave its derivation for a future work.
Instead, we will try now to derive the function $R\left( \omega \right) $
neglecting this term. In other words, we consider energies $\omega $
exceeding $t_{E}^{-1}$ whatever it is.

In Ref. \cite{asaa}, an additional term was added as a regularizer, which
had to be put to zero at the end of the calculations. However, proceeding in
this way the authors of Ref. \cite{asaa} got a result that did not agree
with the one obtained from the Gutzwiller trace formula\cite{bk}. The
discrepancy has been called  ``repetition problem'' and was discussed in a
number of works \cite{aos,gm}.

We want to show now that the result obtained with the ballistic $\sigma $%
-model containing only the term ${\cal F}_{kin}[Q_{{\bf n}}({\bf r})]$, Eq. (%
\ref{e22}), agrees with what one can expect from the trace formulae. In
contrast, the perturbative approach of Ref. \cite{asaa} is not accurate in
this limit.

We start our discussion with Eqs. (\ref{e21}, \ref{e22}), where only the
term ${\cal F}_{kin}[Q_{{\bf n}}({\bf r})]$ is left. The level-level
correlation function $R\left( \omega \right) $, Eq. (\ref{e7}), can be
written as 
\begin{equation}
R\left( \omega \right) =1-%
\mathop{\rm Re}%
I\left( \omega \right) ,  \label{c1a}
\end{equation}

\begin{eqnarray}
I\left( \omega \right)  &=&\frac{1}{2V^{2}}\int \langle \left( Q_{{\bf n}%
}^{44}\left( {\bf r}\right) -1\right)   \nonumber \\
&&\times (Q_{{\bf n}^{\prime }}^{88}\left( {\bf r}^{\prime }\right)
+1)\rangle _{kin}d{\bf r}d{\bf r}^{\prime }d{\bf n}d{\bf n}^{\prime }
\label{c2}
\end{eqnarray}
where we use Eqs.(\ref{c5}), (\ref{c6}) and introduce a notation 
\[
<...>_{kin}=\int \left( ...\right) \exp \left( -{\cal F}_{kin}^{\left(
0\right) }[Q_{{\bf n}}\left( {\bf r}\right) ]\right) DQ_{{\bf n}}
\]

The free energy functional ${\cal F}_{kin}^{\left( 0\right) }$ is obtained
from ${\cal F}_{kin}$, Eq.(\ref{e22}) by putting $\hat{a}=0$.

Due to the absence of any regularizer a perturbation expansion in ballistic
excitations (diffusion modes) cannot be good because one obtains diverging
integrals with propagators like the one in Eq. (\ref{a100}) in the
integrand. Therefore, this method should not be applied and we should try
something different. At the same time, the specific form of the functional $%
{\cal F}_{kin}^{(0)}[Q_{{\bf n}}\left( {\bf r}\right) ]$ that does not
contain second space derivatives allows us to simplify the functional
integral by reducing it to functional integrals on periodic orbits.

In order to proceed in this way we discretize the phase space writing the
functional integral as a definite integral over $Q$ at all sites of a
lattice in the phase space. In this way we write the function $I\left(
\omega \right) $, Eq. (\ref{c2}), as 
\begin{eqnarray}
&&I\left( \omega \right) =\frac{[\Delta \Omega ]^{2}f^2}{2V^{2}}\sum_{{\bf r}%
_{i},{\bf r}_{j},{\bf n}_{i},{\bf n}_{j}}\langle \big(Q_{{\bf n}%
_{i}}^{44}\left( {\bf r}_{i}\right) -1\big)  \nonumber \\
&&\times \big(Q_{{\bf n}_{j}}^{88}\left( {\bf r}_{j}\right) +1\big)\rangle
_{kin}  \label{c3}
\end{eqnarray}

The free energy functional ${\cal F}_{kin}^{\left( 0\right) }$ on this
lattice takes the form 
\begin{equation}
{\cal F}_{kin}^{\left( 0\right) }[Q_{{\bf n}}\left( {\bf r}\right) ]=\sum_{%
{\bf n}_{i}}F_{{\bf n}_{i}}[Q]  \label{c4}
\end{equation}

\begin{eqnarray*}
F_{{\bf n}}[Q] &=&\frac{\pi \nu \lbrack \Delta \Omega ]f}{4}Str[\sum_{\{{\bf %
r_{i}},{\bf r_{i}^{\prime }}\}}\frac{2v_{F}}{f}\Lambda \overline{T}_{{\bf n}%
}\left( {\bf r}_{i}\right) T_{{\bf n}}\left( {\bf r}_{i}^{\prime }\right) \\
&&+\sum_{{\bf r}_{i}}i(\omega +i\delta )\Lambda Q_{{\bf n}}\left( {\bf r}%
_{i}\right) ]
\end{eqnarray*}

In Eqs.(\ref{c3}, \ref{c4}), $f$ is the elementary length in the coordinate
space along a trajectory and $[\Delta \Omega ]$ is the elementary phase
volume in the phase space perpendicular to it (i.e. to the unit vector ${\bf %
n}$). The summation in the first term in $F_{{\bf n}}[Q]$ is performed over
nearest neighbors on the trajectory and in a certain order. The elementary
volume $[\Delta \Omega ]$ in Eqs. (\ref{c3}, \ref{c4}) can be written as 
\begin{equation}
\lbrack \Delta \Omega ]=S_{d}^{-1}\prod_{i=1}^{d-1}\left( \Delta n_{\perp
}^{i}\Delta r_{\perp }^{i}\right)   \label{c4a}
\end{equation}
where $S_{d}$ is the surface of the unit sphere in the $d$-dimensional
space. In Eq. (\ref{c4a}), $\Delta r_{\perp }^{i}$ and $\Delta n_{\perp }^{i}
$ are the elementary length and momentum in a direction perpendicular to the
path.

In principle, the length $f$ and the space volume $[\Delta \Omega ]$ may be
arbitrary. At the same time, we should remember that we have used the
quasiclassical approximation and the length $f$ may not be smaller than the
wavelength $\lambda _{F}$. However, only for specific choice of $[\Delta
\Omega ]$, the functional $F_{{\bf n}}[Q]$, Eq. (\ref{c4}), remains single
valued. As we will see, this choice corresponds to the Bohr-Sommerfeld
quantization rules.

In order to reduce the multiple integral over all $Q$ on the lattice sites
in the phase space to a simpler form we use the following equalities that
can be proven using the methods of integration over supermatrices\cite
{efetov} 
\begin{equation}
\int \exp \left( -F_{{\bf n}}[Q]\right) DQ=1  \label{c5}
\end{equation}
\begin{eqnarray}
&&\int \left( Q_{{\bf n}}^{44}\left( {\bf r}\right) -1\right) \exp \left(
-F_{{\bf n}}[Q]\right) DQ  \label{c6} \\
&=&\int \left( Q_{{\bf n}}^{88}\left( {\bf r}\right) +1\right) \exp \left(
-F_{{\bf n}}[Q]\right) DQ=0  \nonumber
\end{eqnarray}
for any ${\bf r}$ and ${\bf n}$.

Let us understand first how to simplify the function $I\left( \omega \right) 
$, Eq. (\ref{c2}), for an infinite sample. Using Eqs. (\ref{c5}, \ref{c6})
we conclude immediately that only the terms with ${\bf n}_{i}={\bf n}_{j}$
contribute in the sum in Eq. (\ref{c3}). As concerns the free energy
functional, one can integrate over all $Q_{{\bf n}}$ with ${\bf n}_j\neq{\bf %
n}_{i}$ using Eq. (\ref{c5}), which leaves only one term $F_{{\bf n}_{i}}[Q]$
in the exponential. Only the term with ${\bf n}_{i}$ parallel to the line
connecting the points ${\bf r}$ and ${\bf r}^{\prime }$ gives a non-zero
contribution and, moreover, integration over $Q$ on sites outside the line
gives $1$ because the free energy functional ${\cal F}_{kin}^{\left(0\right)
}[Q_{{\bf n}}\left( {\bf r}\right) ]$, Eq. (\ref{c4}), does not contain
couplings of $Q$ on these sites with $Q$ on sites on the line between ${\bf r%
}$ and ${\bf r}^{\prime }$.

Thus, we come to an integral over all $Q$ on sites along the line connecting
the points ${\bf r}$ and ${\bf r}^{\prime }$%
\begin{equation}
I\left( \omega \right) =I_{+}\left( \omega \right) +I_{-}\left( \omega
\right) ,  \label{c7}
\end{equation}

\begin{eqnarray*}
I_{\pm }\left( \omega \right) &=&\frac{[\Delta \Omega ]^{2}f^{2}}{2V^{2}}%
\sum_{{\bf r_{i}},{\bf r_{j}}}\int \left( Q_{\pm }^{44}\left( {\bf r_{i}}%
\right) -1\right) \left( Q_{\pm }^{88}\left( {\bf r_{j}}\right) +1\right) \\
&&\times \exp \left( -F_{+}[Q]-F_{-}[Q]\right) DQ,
\end{eqnarray*}
\begin{eqnarray*}
&&F_{\pm }[Q_{\pm }]=\frac{\pi \nu \lbrack \Delta \Omega ]f}{4}Str[\sum_{\{%
{\bf r_{i}},{\bf r_{i}^{\prime }}\}_{\pm }}\frac{2v_{F}}{f}\Lambda \overline{%
T}_{\pm }\left({\bf r_{i}}\right) T_{\pm }\left({\bf r_{i}^{\prime }}\right) \\
&&+\sum_{r_{i}}i(\omega +i\delta )\Lambda Q_{\pm }\left({\bf r_{i}}\right) ]
\end{eqnarray*}
The signs $+$ and $-$ correspond to different directions of the trajectory;
the summation over the pairs $\{{\bf r_{i}},{\bf r_{i}^{\prime }}\}_{\pm }$
in the free energies $F_{\pm }[Q_{\pm }]$ should be fulfilled in the order
conforming with its direction. One can see that the free energies and,
hence, both the terms $I_{+}\left( \omega \right) $, $I_{-}\left( \omega
\right) $ are equal to each other. This relation follows formally from the
definition of the conjugation and the equality $Q_{{\bf n}}({\bf r})=%
\overline{Q}_{{\bf n}}({\bf r})$. It is important that the time-reversal
symmetry is not violated.

We see from Eqs. (\ref{c7}) that the functional integral, Eq. (\ref{c2}),
over $Q_{{\bf n}}\left( {\bf r}\right) $ on all sites in the phase space of
an infinite sample has been reduced to a functional integral along a line
and averaging over all directions of this line. This is due to a specific
form of the free energy functional containing only first space derivatives.
Any regularizer containing second derivatives would make such a reduction
impossible.

What happens if the sample is finite? We can reduce as before the functional
integral, Eq. (\ref{c2}), to an integral over the line. However, we can
follow this line until we reach the boundary, where we have a degeneracy
that follows from the boundary condition, Eq. (\ref{e16c}). Namely, the
supermatrix $Q_{{\bf n}}$ belongs to $2$ different lines. This means that,
having reached the surface, we can follow the line obtained from the first
one by a specular reflection. We can keep going along the second line until
we reach another boundary, etc. In principle, we have two possibilities:

1. After several reflections from the boundaries we come to the same point
in the phase space or, in other words we get a periodic orbit.

2. The broken line obtained after the reflections on the boundaries does not
close in the phase space, which can be considered as a periodic orbit with
an infinite period. Of course, we should speak rather of tubes than of
lines. However, this is not important because we are interested now in
comparatively high frequencies $\omega $ of the order of a typical period of
the orbit. Very long orbits would contribute to the function $I\left( \omega
\right) $, Eq. (\ref{c2}), at much smaller frequencies. The question about
the finite thickness of the lines can arise for times larger than the
Ehrenfest time $t_{E}$, Eq. (\ref{c1}), which we do not consider here.

So, let us assume that we have got a periodic orbit. All supermatrices $T$
and $Q$ $\ $in Eqs. (\ref{c7}) are assumed to be on this orbit and we can
write these equations in a more convenient form choosing the thickness of
the paths and changing to the continuous limit along them. Taking the
continuous limit along the paths the elementary length $f$ will disappear.
The only quantity to be chosen is the elementary phase volume $[\Delta
\Omega ]$ in Eq. (\ref{c4a}).

In the $d$-dimensional space the density of states $\nu $ can be written as 
\begin{equation}
\nu =\frac{dn}{d\varepsilon _F }=\frac{\Omega _{d}d}{\left( 2\pi \hbar
\right) ^{d}}\frac{p_{F}^{d-1}}{v_F}  \label{c8}
\end{equation}
where $\Omega _{d}$ is the volume of the unit $d$-dimensional sphere.

Then, using the relation between the surface of the unit sphere and its
volume $S_{d}=\Omega _{d}d$ we reduce the coefficient $\nu \lbrack \Omega ]$
entering the free energy functional $F_{\pm }[Q]$ in Eqs. (\ref{c7}) to the
form 
\begin{equation}
\nu \lbrack \Delta \Omega ]=\frac{1}{\left( 2\pi \hbar \right) ^{d}v_{F}}%
\prod_{i=1}^{d-1}\left( \Delta p_{\perp }^{i}\Delta r_{\perp }^{i}\right)
\label{c9}
\end{equation}
Now we have to choose the product $\prod_{i=1}^{d-1}\left( \Delta p_{\perp
}^{i}\Delta r_{\perp }^{i}\right) $ and we do this using a standard
quasiclassical rule according to which we write 
\begin{equation}
\prod_{i=1}^{d-1}\left( \Delta p_{\perp }^{i}\Delta r_{\perp }^{i}\right)
=\left( 2\pi \hbar \right) ^{d-1}  \label{c10}
\end{equation}
With Eq. (\ref{c10}) we obtain 
\begin{equation}
\nu \lbrack \Delta \Omega ]=\left( 2\pi \hbar v_{F}\right) ^{-1}  \label{c11}
\end{equation}
Putting as everywhere before $\hbar =1$ we can rewrite Eqs. (\ref{c7}) as a
sum over periodic orbits 
\begin{eqnarray}
I\left( \omega \right) &=&\frac{\Delta ^{2}}{\left( 2\pi v_{F}\right) ^{2}}%
\sum_{p}\int \langle \left( Q^{44}\left( x\right) -1\right)  \nonumber \\
&&\times \left( Q^{88}\left( x^{\prime }\right) +1\right) \rangle
_{1}dxdx^{\prime }  \label{c12}
\end{eqnarray}
where $<...>_{1}$ stands for the functional integral 
\begin{equation}
\langle ...\rangle _{1}=\int \left( ...\right) \exp \left( -F_{1}[Q]\right)
DQ  \label{c12a}
\end{equation}
and $\Delta =\left( \nu V\right) ^{-1}$ is the mean level spacing for the
billiard under consideration. The one-dimensional free energy functional $%
F_{1}\left[ Q\right] $ for an orbit takes the form 
\begin{equation}
F_{1}\left[ Q\right] =\frac{1}{2}Str\int \left( \Lambda \bar{T}\left(
x\right) \frac{dT\left( x\right) }{dx}+\frac{i\left( \omega +i\delta \right) 
}{2v_{F}}\Lambda Q\right) dx  \label{c13}
\end{equation}
and is the result of the adding of the free energies $F_{\pm }[Q_{\pm }]$
Eq.(\ref{c7}). The sum over $p$ in Eq. (\ref{c12}) means the sum over all
periodic orbits and, in principle, $Q\left( x\right) $ should depend on $p$.
In order to simplify notations we omit writing this dependence explicitly.
The integrals over $x$ and $x^{\prime }$ are taken along the orbits (a
certain direction is implied to be already chosen).

The overall coefficient in the functional $F_{1}\left[ Q\right] $ is
determined by the quasiclassical rule, Eq. (\ref{c10}). It is not difficult
to understand that Eq. (\ref{c10}) is the only reasonable choice for the
``thickness'' of the classical paths. The functional $F_{1}\left[ Q\right] $
is multivalued because any replacement of the type $T\left( x\right)
\rightarrow T\left( x\right) h\left( x\right) $, where $\bar{h}\left(
x\right) h\left( x\right) =1$ and $[h\left( x\right) ,\Lambda ]=0$, does not
change the supermatrix $Q\left( x\right) $ but changes the functional $F_{1}%
\left[ Q\right] $ as 
\begin{eqnarray}
F_{1}\left[ Q\right] &\rightarrow &F_{1}\left[ Q\right]  \label{c14} \\
&&+\frac{1}{2}\int Str\left( \bar{h}_{1}\left( x\right) \frac{dh_{1}\left(
x\right) }{dx}-\bar{h}_{2}\left( x\right) \frac{dh_{2}\left( x\right) }{dx}%
\right) dx  \nonumber
\end{eqnarray}
where $h_{1}$ and $h_{2}$ are the upper and lower diagonal blocks,
respectively. Writing $h_{m}$, $m=1,2$ as 
\begin{equation}
h_{m}\left( x\right) =\left( 
\begin{array}{cc}
\exp \left( i\hat{\phi}_{m}\left( x\right) \right) & 0 \\ 
0 & \exp \left( i\hat{\chi}_{m}\left( x\right) \right)
\end{array}
\right) ,  \label{c15}
\end{equation}
\begin{eqnarray*}
\hat{\phi}_{m}\left( x\right) &=&\left( 
\begin{array}{cc}
\phi _{m}^{+}\left( x\right) & 0 \\ 
0 & -\phi _{m}^{-}\left( x\right)
\end{array}
\right) , \\
\text{ }\hat{\chi}_{m}\left( x\right) &=&\left( 
\begin{array}{cc}
\chi _{m}^{+}\left( x\right) & 0 \\ 
0 & -\chi _{m}^{-}\left( x\right)
\end{array}
\right)
\end{eqnarray*}
we write the change $\Delta F_{1}\left[ Q\right] $ of the functional $F_{1}%
\left[ Q\right] $ as 
\begin{equation}
\Delta F_{1}\left[ Q\right] =\frac{i}{2}\sum_{m=1,2}\left( -1\right)
^{m-1}\left( \Delta \phi _{m}^{+}-\Delta \phi _{m}^{-}+\Delta \chi
_{m}^{+}-\Delta \chi _{m}^{-}\right)  \label{c16}
\end{equation}
where $\Delta \phi _{m}^{\pm }$ and \ $\Delta \chi _{m}^{\pm }$ are changes
of the phases when going around the periodic orbit. These changes must be
integer multiple of $2\pi .$ The phases $\phi _{m}^{+}\left( x\right) $ and $%
\phi _{m}^{-}\left( x\right) $ (as well as $\chi _{m}^{+}\left( x\right) $
and $\chi _{m}^{-}\left( x\right) $) are not independent of each other. If $%
\phi _{m}^{+}\left( x\right) =\left( 2\pi k_{m}/L_{p}\right) x$, then $\phi
_{m}^{-}\left( x\right) =-\left( 2\pi k_{m}/L_{p}\right) x$ ($k_{m}$ is an
integer and $L_{p}$ is the length of a $p$-orbit). In a general case we can
write 
\begin{equation}
\phi _{m}^{\pm }\left( x\right) =\pm \frac{2\pi k_{m}}{L}x\tilde{\phi}%
_{m}^{\pm }\left( x\right) ,\text{ \ }\tilde{\phi}_{m}^{\pm }\left( 0\right)
={\bf \tilde{\phi}}_{m}^{\pm }\left( L_{p}\right)  \label{c17}
\end{equation}
and the same for $\chi _{m}^{\pm }$. Then, we obtain for the changes of the
phases $\Delta \phi _{m}^{+}=-\Delta \phi _{m}^{-}$, \ $\Delta \chi
_{m}^{+}=-\Delta \chi _{m}^{-}$, which reduces Eq. (\ref{c16}) to the form 
\begin{equation}
\Delta F_{1}\left[ Q\right] =2\pi iM  \label{c18}
\end{equation}
where $M$ is integer.

With Eq. (\ref{c18}) we come to the result that, although the functional $%
F_{1}\left[ Q\right] $ is multivalued, the partition function $\exp \left(
-F_{1}\left[ Q\right] \right) $ is not and one can integrate over $Q$ in a
standard way. In principle, we could proceed in the opposite way and
determine the thickness of the classical paths by demanding the partition
function be single valued. Then, we would obtain the quasiclassical
quantization rule, Eq. (\ref{c10}), automatically.

Now, let us calculate the functional integral, Eq. (\ref{c12}), using two
different approaches. First, we compute the functional integral following
the scheme of Ref.\cite{asaa}. We can use different parametrizations like
those specified by Eqs. (\ref{e23}) or (\ref{a23}). The parametrization, Eq.
(\ref{a23}), is very convenient because the Jacobian is equal to unity. At
the same time, the ballistic $\sigma $-model contains also non-harmonic in $%
P $ terms. In the quadratic approximation, one writes the free energy as 
\begin{equation}
F_{1}^{\left( 2\right) }[Q]=\frac{1}{2}Str\int \left( \Lambda P\left(
x\right) \frac{dP\left( x\right) }{dx}-\frac{i\left( \omega +i\delta \right) 
}{v_{F}}P^{2}\left( x\right) \right) dx  \label{c19}
\end{equation}
The supermatrices $Q\left( x\right) $ entering Eq. (\ref{c12}) should also
be expanded up to quadratic terms. Calculating gaussian integrals we obtain
for $I\left( \omega \right) $%
\begin{equation}
I\left( \omega \right) =\frac{\Delta ^{2}}{\pi ^{2}v_{F}^{2}}%
\sum_{p}\sum_{m=-\infty }^{\infty }\left( q_{m}^{\left( p\right) }+\frac{%
\omega +i\delta }{v_{F}}\right) ^{-2},  \label{c20}
\end{equation}
where 
\[
q_{m}^{\left( p\right) }=\frac{2\pi m}{L_{p}} 
\]
are momenta corresponding to a $p$ -periodic orbit, $L_{p}$ is the length of
the orbit.

Eq. (\ref{c20}) corresponds to the ``perturbative part'' of the level-level
correlation function of Ref. \cite{asaa}, although here we sum over momenta
on periodic orbits instead of summation over eigenvalues of the
Perron-Frobenius operator. However, it is clear that Eq. (\ref{c20}) cannot
correspond to a good perturbation theory because this expression contains
resonances at arbitrarily high frequencies.

One might guess that next orders of the expansion in $P$ had to be taken
into account and the expansion would not be good at the resonances.
Curiously enough, it is not so. One can see immediately that the quadratic
form, Eq. (\ref{c19}) is ${\em exact}$ in the parametrization, Eq. (\ref{e23}%
) (the overall coefficient is $2$ times larger). At the same time, the part $%
Q^{\parallel \text{ }}$of the supermatrix $Q$ commuting with $\Lambda $ and
entering Eq. (\ref{c12}) is {\em exactly} $Q^{\parallel }=1+2P^{2}$.
Therefore, we do not obtain any perturbative corrections to Eq. (\ref{c20})
\ (at the same time, the contribution of the Jacobian in the
parametrization, Eq. (\ref{e23}), is not as clear).

Nevertheless, Eq. (\ref{c20}) is not exact. The matrices $P$ in Eq. (\ref
{e23}) vary on non-trivial manifolds and extending the integration over
these matrices from $-\infty $ to $+\infty $ as it is implied in any
gaussian integration is unjustified. To make the discussion simpler, let us
rewrite Eqs. (\ref{c1a}, \ref{c20}) using the Poisson summation formula as 
\begin{equation}
R\left( \omega \right) =1+\sum_{p}\left( \frac{T_{p}\Delta }{\pi }\right)
^{2}%
\mathop{\rm Re}%
\sum_{n=1}^{\infty }n\exp \left( i\left( \omega +i\delta \right)
T_{p}n\right)   \label{c21}
\end{equation}
where $T_{p}=L_{p}/v_{F}$ is the period of the motion on the $p$ -orbit. Eq.
(\ref{c21}) corresponds to an expansion in periodic orbits of a classical
flow \cite{cvit}. Strictly speaking, Eq. (\ref{c21}) is different from what
one writes for classical flows by absence of a factor containing the
monodromy matrix. It is clear that in our simple consideration the stability
of the periodic orbit is not taken into consideration. As any periodic orbit
we consider has a finite thickness, the monodromy matrix would appear in a
more accurate calculation. However, the aim of this chapter is only to
clarify the origin of the repetition problem and therefore we use the
simplest approximation.

The factor $n$ in front of the exponential is a characteristic feature of 
expansions for classical flows (see e.g. Eq. (52) of Ref.\cite{cvit} which leads to this
dependence after taking the logarithm of both parts and taking second
derivative in $s$). In other words, we have now an expansion in periodic
orbits of the Perron-Frobenius operator and this corresponds to the result
of Ref. \cite{asaa} in the limit of the vanishing regularizer.

However, although the perturbative approximation of Ref. \cite{asaa} works
very well in the diffusion limit, we do not see any justification for it in
the ballistic limit. Therefore, we should try to calculate the integral in
Eqs. (\ref{c12}, \ref{c12a}) without using this approximation.

Fortunately, the functional integral in Eq. (\ref{c12}, \ref{c12a}) can be
calculated exactly even easier than approximately. The free energy
functional $F_{1}[Q]$ entering these equations corresponds to a
one-dimensional ring without any impurities, provided the averaging over the
spectrum has been performed. This energy averaging is necessary to get a
smooth quasiclassical function $g_{{\bf n}}\left( {\bf r}\right) $, Eq. (\ref
{klassaver}). Without the averaging this function would not be smooth due to
quantization of the energy levels in the ring. So, we conclude that the
calculation of functional integral over $Q$ with the free energy functional $%
F_{1}[Q]$ is equivalent to calculation of the averaged level-level
correlation function $R_{p}\left( \omega \right) $ for a clean electron
system on a ring.

The level-level correlation function $R_{1p}\left( \omega \right) $ for such
a ring can be written as 
\begin{eqnarray}
R_{1p}\left( \omega \right) =\left( \nu _{1}L_{p}\right)
^{-2}\sum_{m,m^{\prime }=-\infty}^{\infty } &&\langle \delta \left( \varepsilon
-\omega -\varepsilon \left( q_{m}\right) \right)  \nonumber \\
\times &&\delta \left( \varepsilon -\varepsilon \left( q_{m^{\prime
}}\right) \right) \rangle _{\varepsilon }  \label{c22}
\end{eqnarray}
where $\langle\dots\rangle_{\varepsilon }$ means the averaging over $\varepsilon $, $%
q_{m}=2\pi m/L_{p}$ and $\nu _{1}=\left( \pi v_{F}\right) ^{-1}$ is
one-dimensional density of states. The spectrum $\varepsilon \left( q\right) 
$ can be, as usual, linearized 
\begin{equation}
\varepsilon \left( q_{m}\right) =\frac{q_m^{2}-p_{F}^{2}}{2m}\approx
v_{F}\left( |q_{m}|-p_{F}^{\left( p\right) }\right)  \label{c23}
\end{equation}

Using the Poisson formula and Eq. (\ref{c23}) we write 
\begin{equation}
\sum_{m=0}^{\infty }\delta \left( \varepsilon -\varepsilon \left(
q_{m}\right) \right) \approx \frac{L_p}{\pi v_{F}}\sum_{n=-\infty }^{\infty
}\exp \left( i\left( 2\pi n_{F}^{\left( p\right) }+\varepsilon T_{p}\right)n
\right)  \label{c24}
\end{equation}
where $n_{F}^{\left( p\right) }=p_{F}^{\left( p\right) }L_{p}/2\pi $.

Substituting Eq. (\ref{c24}) into Eq. (\ref{c22}) we obtain 
\begin{equation}
R_{1p}\left( \omega \right) =\sum_{n,n^{\prime }=-\infty }^{\infty }\exp
\left( i(2\pi n_{F}^{\left( p\right) }+\varepsilon T_{p})(n-n^{\prime
})+\omega T_{p}n^{\prime }\right)   \label{c25}
\end{equation}
After averaging over $\varepsilon $, only the term with $n=n^{\prime }$
gives the contribution in Eq. (\ref{c25}). On the other hand, repeating all
the steps of the derivation of the $\sigma $-model, we come to Eqs. (\ref
{c12}, \ref{c12a}) with the only difference that we should replace the mean
level spacing $\Delta $ of the entire system by the level spacing of the
one-dimensional ring $\Delta _{p}=\pi /T_{p}$. This allows us to write the
level-level correlation function $R\left( \omega \right) $ of the quantum
billiard under consideration as 
\begin{equation}
R\left( \omega \right) =1+\sum_{p}\left( \frac{T_{p}\Delta }{\pi }\right)
^{2}%
\mathop{\rm Re}%
\sum_{n=1}^{\infty }\exp \left( i\left( \omega +i\delta \right)
T_{p}n\right)   \label{c26}
\end{equation}
Comparing Eqs. (\ref{c21}) and (\ref{c26}) with each other we see that the
only difference between them is the presence of the prefactor $n$ in Eq. (%
\ref{c21}). So, we conclude that the factor $n$ in the expansion in periodic
orbits is a consequence of a unjustified approximation of Ref. \cite{asaa}
and this solves the problem of repetitions \cite{bk}. The assumption of Ref. 
\cite{asaa} that the regularizer can be put to zero at the end of
calculations does not seem to be correct. We believe (following Ref. \cite
{al}) that the presence of a finite regularizer is inevitable in a quantum
system and a very important problem is to calculate it.

The problem of repetitions was discussed recently for weak scatterers in
Ref. \cite{gm} where the problem was related to the question of a
possibility of separating 4 point correlation functions into two diffusons. 
From the above discussion, we see that the problem is even more delicate
because Eq. (\ref{c20}) is perturbatively exact and the difference comes
from oscillating exponentials.

In order to understand better what has been neglected in our derivation, we
compare Eq. (\ref{c26}) with a corresponding diagonal contribution obtained
from the Gutzwiller trace formula, (see, e.g. \cite{keat}) 
\begin{equation}
R^{\left( d\right) }\left( x\right) =1+\frac{2}{T_{H}^{2}}%
\sum_{p}\sum_{n=1}^{\infty }g_{p}\frac{T_{p}^{2}}{\left| M_{p}^{n}-I\right| }%
\cos \left( \frac{2\pi nT_{p}}{T_{H}}x\right)  \label{c27}
\end{equation}
where $g_{p}$ is the action-multiplicity of the $p^{th}$ primitive orbit
(for the orthogonal ensemble $g_{p}=2$), and $M_{p}$ is the monodromy matrix
that describes the flow linearized in its vicinity. In Eq. (\ref{c27}), $%
T_{H}=2\pi /\Delta $ is the Heisenberg time, $T_{p}$ is the orbit period,
and $x=\omega /\Delta $. Except for the factor $\left| M_{p}^{n}-I\right|
^{-1}$ Eqs. (\ref{c26}) and (\ref{c27}) agree. As we carried out computation
without any regularizer like the one of Ref. \cite{al}, we conclude again
that its presence is absolutely necessary and it must be related to the
monodromy matrix $M_{p}$. We see that without this term correlations between
orbits do not exist and one cannot pass to the universal limit when lowering
the frequency $\omega $. In the language of the field theoretical approach,
one cannot reduce the ballistic $\sigma $-model to the zero-dimensional one
without this regularizing term that must describe quantum diffraction on
irregularities of the boundary.

Adding a term like 
\begin{equation}
F_{reg}\left[ Q\right] =Str\int \beta _{{\bf n}}\left( {\bf r}\right) \left( 
\frac{\partial Q\left( {\bf r}\right) }{\partial {\bf n}}\right) ^{2}d{\bf r}%
d{\bf n}  \label{c28}
\end{equation}
where $\beta _{{\bf n}}\left( {\bf r}\right) $ is a function in the phase
space, we may obtain effectively a coupling between the periodic orbits of
the type 
\begin{equation}
-\sum_{p,p^{\prime }}\beta _{pp^{\prime }}Str\int Q^{\left( p\right) }\left(
x_{p}\right) Q^{\left( p^{\prime }\right) }\left( x_{p^{\prime }}\right)
dx_{p}dx_{p^{\prime }}  \label{c29}
\end{equation}
which resembles coupling between grains in a granular system. For small $%
\beta _{p,p^{\prime }}$ the orbits are not coupled and we have separate
periodic orbits. As the coupling $\beta _{p,p^{\prime }}$ grows, the
relative fluctuations of $Q^{\left( p\right) }$ with respect to each other
get suppressed and one needs to consider rotations of the system as the
whole. Then, one obtains the zero-dimensional $\sigma $-model and, hence,
the Wigner-Dyson statistics. This is a scenario for a granular metal and we
believe that it is relevant for the quantum billiard, the role of the grains
being played by separate periodic orbits.

\section{ Discussion.}

In the present work, we made an attempt to put the field theoretical
approach to systems with a long range disorder on a solid basis. The
conventional method of derivation of the supermatrix $\sigma $-model \cite
{efetov} is based on singling out slow modes, performing
Hubbard-Stratonovich transformation and using a saddle-point approximation.
Although this approach worked well for a short range disorder, its validity
is not justified for a long range disorder and quantum billiards. We
suggested a scheme that allows us to overcome these difficulties and derive
a modified non-linear ballistic $\sigma $-model (see Eq.(\ref{e22})).

The method resembles the approach of Ref. \cite{mk} and is based on writing
quasiclassical equations for generalized Green functions. At the same time,
the quasiclassical equations are written for non-averaged over the long
range potential quantities and singling out slow modes is performed only for
a part originating from a short range disorder. In addition, the short range
disorder does not play an important role and can be put to zero. The crucial
step of the derivation is that the solution of the quasiclassical equations
can be found exactly, which is a consequence of the supersymmetric structure
of the Green functions. This possibility was overlooked in the previous
study \cite{mk}.

The scheme developed now leads to a considerable progress in describing
disordered systems with long range disorder because the derivation is
applicable for all lengths exceeding the wavelength $\lambda _{F}$.
Representing the solution of the quasiclassical equations and also the
partition function for the electron Lagrangian with sources in terms of a
functional integral over supermatrices $Q_{{\bf n}}\left( {\bf r}\right) $, $%
{\bf n}^{2}=1$, with the constraint $Q_{{\bf n}}^{2}\left( {\bf r}\right) =1$
we were able to average over the disorder exactly and obtain a ballistic $%
\sigma $-model in a new form that has not been written before. The so called
``mode locking'' \cite{aos} problem does not arise here because the
eigenvalues of the supermatrix $Q_{{\bf n}}\left( {\bf r}\right) $ are fixed
by the construction and do not fluctuate. The method suggested resembles the
method of bosonization, well known in field theory, see e.g. a book \cite
{gnt}, when a fermionic system is replaced by a bosonic one. In our
approach, we also replace the electron system by a system of ballistic
excitations that can be considered as quasiparticles. At large scales, these
quasiparticles are well known diffusons and cooperons. In analogy, our
scheme can be called superbosonization.

For weak scatterers, there should exist $2$ more scales: the Lapunov length $%
l_{L}$ and the transport mean free path $l_{tr}$. The single particle mean
free path $l$ does not appear in our consideration (actually, we do not
consider one-particle Green functions at different points restricting our
study to gauge invariant quantities). Integrating out degrees of freedom
related to distances smaller than the Lapunov length $l_{L}$ we obtained a
reduced ballistic $\sigma $-model. A propagator describing small
fluctuations within this reduced $\sigma $-model corresponds to the kinetic
Boltzmann equation with a collision term. Integrating further on scales up
to the transport mean free path $l_{tr}$ we obtain the standard diffusive $%
\sigma $-model.

Trying different calculational schemes we conclude that one can do
perturbative calculations with the ballistic $\sigma $-model only at scales
exceeding the Lapunov length (we call this range ``collision region''). At
smaller lengths (following Ref. \cite{al} we call this range Lapunov region)
no perturbation expansions in diffusons and cooperons are possible. In this
region one can carry out calculations deriving equations for correlation
functions and investigating them in different approximations.

It seems that an infinite system with a weak long range disorder is
adequately described by the ballistic $\sigma $-model we have derived.
However, when describing quantum billiards, our approach is not accurate
near the boundaries, where the quasiclassical approximation may not be used
(turning points). We believe that a more accurate derivation may result in a
new term in the $\sigma $-model. This term was suggested phenomenologically
in Refs. \cite{al,al1} but has not been derived yet microscopically. Its
presence seems to be absolutely necessary because it must introduce a new
scale: the Ehrenfest time $t_{E}$. One may not put the regularizer to zero
at the end of calculations. We have demonstrated that neglecting such a term
we reduced the ballistic $\sigma $-model for the billiard to ballistic $%
\sigma $-models for periodic orbits. Proceeding in this way we demonstrated
explicitly where the contradiction between the work\cite{asaa} and Ref. \cite
{bk}(repetition problem) comes from. Our conclusion is that the
representation of the level-level correlation function in terms of
eigenvalues of the Perron-Frobenius operator suggested in Ref. \cite{asaa}
is not justified in the ballistic case. This approach is valid only if there
are no repetitions but this would rather correspond the diffusive case. In
the opposite limit, one comes to a description in terms of periodic orbits
without correlations between actions of different orbits. This is the region
where the description of Ref. \cite{bk} may be applicable. At the same time,
the Ehrenfest time can hardly be identified on the basis of the trace
formula and therefore the limits of applicability of the result of Ref.\cite
{bk} have not been specified. The hypothetical regularizer seems to be
related to the monodromy matrix entering the Gutzwiller trace formula.

We believe that the field theoretical approach presented here and the
formalism based on the Gutzwiller trace formula can be complementary to each
other describing quantum systems in different regions of parameters. At
times smaller than the Ehrenfest time, the trace formula can be more
convenient. However, at a larger time, trying to extract physical quantities
from the trace formula does not make much sense because the $\sigma $-model
is a much more convenient tool for such calculations. This concerns
especially the universal limit where the $\sigma $-model approach leads for
most correlation functions to definite integrals that can be computed rather
easily.

It is important to notice, that diagrammatic expansions like those attempted
in Ref. \cite{lerner} can hardly be successful. The authors of Ref. \cite
{lerner} found that the results depended crucially on the way how the
ultraviolet cutoff was introduced. Now we understand that the ultraviolet
cutoff must be imposed by the requirement that the integration is performed
over the manifold $Q_{{\bf n}}^{2}\left( {\bf r}\right) =1$ in an invariant
way. Any artificial ultraviolet cutoffs in the perturbation theory would
correspond to a violation the rotational invariance in the space of the
supermatrices $Q_{{\bf n}}\left( {\bf r}\right) $ and lead to wrong results.
At the same time, it is not clear how to develop a perturbation theory in an
invariant way.

An important question of an averaging procedure was discussed in several
works \cite{zirn,aos,bmm,gm}. In all these publications an opinion was
expressed that an averaging over energy was not sufficient for study of
quantum chaos in quantum systems and different types of an additional
averaging were suggested. We do not agree with this point of view because,
in our derivation of the quasiclassical equations, averaging over the energy
allowed us to smooth generalized Green functions and this was all we needed.
The only condition is that the averaging should be performed in an interval
of energies much exceeding the mean level spacing.

The source of the discrepancy is simple: the authors of the works \cite
{aos,bmm,gm} used the saddle point approximation and the expansion in
gradients. Although the saddle point approximation was not necessary in Ref. 
\cite{zirn}, the gradient expansion still had to be carried out. Therefore,
an additional averaging was necessary to justify these approximation. Since
we do not do such approximations, no additional averaging is needed in our
scheme.

In conclusion, our approach enables us to carry out calculations for long
range disorder and chaos in a reliable way. Still, a derivation of a new
term (\ref{c28}) in the $\sigma $-model (\ref{e22}) describing quantum
diffraction on boundaries has to be done to make the theory complete but we
believe that this is not impossible.

\ 

\section{Acknowledgments}

We acknowledge the financial support of the Sonderforschungsbereich 237 and
GRK 384. A part of the work is a result of participation of one of the
authors (K.B.E.) in the program ``Chaos and Interactions: from Nuclei to
Quantum Dots'' in the Institute for Nuclear Theory of the University of
Washington. Very useful discussions with O. Agam, I. Aleiner, B.L.
Altshuler, O. Bohigas, A.I. Larkin and I.V. Lerner concerning the present
work are greatly appreciated.

\appendix

\section{%
\index{appendix}Boundary conditions}

In this Appendix, we derive boundary conditions for the boundary of the
sample. We describe the boundary by an external field $u_{B}({\bf r})$ which
is negligible inside the sample and grows sharply at the surface of the
sample. In such a situation, the electron wave function decays fast outside
the sample and, in the limit of infinite potential walls, one can just put
the wave function equal to zero at the boundary. Unfortunately, such a
boundary condition is not very helpful because in the quasiclassical
approximation we use it looses its validity at a distance of several wave
lengths from the boundary and matching the wave functions in the bulk and at
the boundary is necessary.

In order to find effective boundary conditions for quasiclassical Green
functions we follow methods well developed in superconductivity theory \cite
{zaitsev}. First, we write the supermatrix $G^{\alpha \beta }({\bf r};{\bf %
r^{\prime }})$ in a form of a sum over eigensupervectors $\psi _{k}({\bf r})$

\begin{equation}
G^{\alpha \beta }({\bf r};{\bf r^{\prime }})=i\sum_{k}%
\frac{\psi _{k}^{\alpha }({\bf r}){\bar{\psi}}_{k}^{\beta }({\bf r^{\prime }}%
)}{\varepsilon _{k}-\varepsilon }  \label{sum}
\end{equation}
satisfying the Schr\"{o}dinger equation 
\begin{equation}
\left[ H_{0{\bf r}}+u\left( {\bf r}\right) +\Lambda \frac{\omega +i\delta }{2%
}+iJ\left( {\bf r}\right) \right] \psi _{k}\left( {\bf r}\right)
=\varepsilon _{k}\psi _{k}\left( {\bf r}\right)  \label{eigen}
\end{equation}
We assume that the potential $u\left( {\bf r}\right) $ in Eq. (\ref{eigen})
contains not only the impurity field but also the potential $u_{B}\left( 
{\bf r}\right) $ describing the boundary. The conjugated equation can be
written as 
\begin{equation}
{\bar{\psi}}_{k}({\bf r})\left[ H_{0{\bf r}}+u\left( {\bf r}\right) +\Lambda 
\frac{\omega +i\delta }{2}+iJ\left( {\bf r}\right) \right] =\varepsilon _{k}{%
\bar{\psi}}_{k}({\bf r})  \label{conj}
\end{equation}

The summation in Eq.(\ref{sum}) should be performed over the complete set of
eigenfuctions, so that $\sum_{k}\psi _{k}^{\alpha }({\bf r}){\bar{\psi}}%
_{k}^{\beta }({\bf r^{\prime }})=\delta ^{\alpha \beta }\delta ({\bf r}-{\bf %
r^{\prime }})$. It means that the choice of the set of eigenfunctions $\psi
_{k}^{\alpha }({\bf r})$ and the operation of the conjugation must conform
with each other. For example, at distances from the boundary much larger
than the wavelength $\lambda _{F}$ but much smaller than the radius $b$ of
the random potential, we choose the eigenfunctions $\psi _{k}^{\alpha }({\bf %
r})$ in a form of plane waves 
\begin{equation}
\psi _{{\bf p}}({\bf r})=e^{i{\bf p}{\bf r}}\psi _{{\bf p}},  \label{ap1}
\end{equation}
where $\psi _{{\bf p}}$ is a normalized vector from the superspace. Then, we
have to adjust the definition of the conjugation written in the book \cite
{efetov} and add to it the momentum inversion ${\bf p}\rightarrow -{\bf p}$.
If the motion cannot be treated as free changing the sign of the momentum
has to be generalized by replacing this operation by the time reversal.
Since the energy $\varepsilon _{k}$ in Eq.(\ref{eigen}) remains the same
after the time reversal, the spectral expansion written in Eq.(\ref{sum}) is
in agreement with Eq.(\ref{e14a}). Below, we will use this relation between
supervectors $\psi ^{\alpha }({\bf r})$ and those conjugated to them.

Now we introduce local coordinates $(z,{\bf r}_{\shortparallel })$ in the
vicinity of the boundary. The coordinate ${\bf r}_{\shortparallel }$ is a
coordinate along the boundary surface and $z$- is a distance between a given
point and the surface. Points on the surface have the coordinates $(0,{\bf r}%
_{\shortparallel })$. If the boundary is rough, the coordinate system $(z,%
{\bf r}_{\shortparallel })$ is not very useful. However, if the boundary is
smooth, which means that the derivative of the field $u_{B}({\bf r})$ along
it is small in comparison to that in the perpendicular direction, the
coordinates $\left( z,{\bf r}_{\shortparallel }\right) $ are very convenient
for the quasiclassical approximation.

If the radius of the curvature of the boundary is large the electron wave
function in the vicinity of the boundary can be represented as a sum of
one-dimensional solutions with respect to the $z$-direction with amplitudes
slowly dependent on the coordinates ${\bf r}_{\shortparallel }$. Since we
need to know the wave function in the domain in which the potential $u_{B}(%
{\bf r})$ vanishes, we can write the asymptotic form of the wave functions as

\begin{equation}
\psi _{k}({\bf r})=e^{i{\bf p}_{\shortparallel }{\bf r}_{\shortparallel
}}\left( e^{ip_{z}z}\varphi _{k,c}({\bf r}_{\shortparallel
})+e^{-ip_{z}z}\varphi _{k,r}({\bf r}_{\shortparallel })\right)
\label{asympt}
\end{equation}
where $\varphi _{k,c}({\bf r}_{\shortparallel })$ and $\varphi _{k,r}({\bf r}%
_{\shortparallel })$ are some slowly varying supervector functions depending
on the potential $u_{B}\left( {\bf r}\right) $ that play the role of
amplitudes of the coming and reflecting waves respectively. They comprise
the minimal knowledge about the potential $u_{B}({\bf r})$ that is needed to
find the required boundary conditions.

If the boundary is impenetrable and the potential reflects all waves, then
the amplitudes $\varphi _{k,c}({\bf r}_{\shortparallel })$, $\varphi _{k,r}(%
{\bf r}_{\shortparallel })$ can be determined from the condition that the
component of the current perpendicular to it is equal to zero. This
condition is valid not only in the region in which the potential $u_{B}({\bf %
r})$ is relevant but also in the quasiclassical region because the current
cannot considerably change at distances of the order of the wavelength.

An expression for the current can be obtained in a standard way from the
particle conservation law that follows from Eqs.(\ref{eigen}), (\ref{conj}).
Its $z$-component perpendicular to the surface is proportional to the
difference $\partial _{z}{\bar{\psi}}\psi -{\bar{\psi}}\partial _{z}\psi $.
Substituting Eq.(\ref{asympt}) and its conjugate into the difference and
putting the result to zero we find ${\bar{\varphi}}_{k,c}({\bf r}%
_{\shortparallel })\varphi _{k,c}({\bf r}_{\shortparallel })={\bar{\varphi}}%
_{k,r}({\bf r}_{\shortparallel })\varphi _{k,r}({\bf r}_{\shortparallel })$.
A relation between the amplitudes of the coming and reflected waves in the
case of the impenetrable boundary can also be established by the demand that
they should transform one into the other by the time reversal. A general
expression that satisfies it can be chosen in the form:

\begin{equation}
\psi _{k}({\bf r})=e^{i{\bf p}_{\shortparallel }{\bf r}_{\shortparallel
}}\left( e^{ip_{z}(z-z_{0})}-e^{-ip_{z}(z-z_{0}}\right) \varphi _{k}({\bf r}%
_{\shortparallel })  \label{wavefunction}
\end{equation}
where $z_{0}$ determines an unknown phase that can be found only by matching
the function $\psi _{k}\left( {\bf r}\right) $, Eq. (\ref{wavefunction}),
with the corresponding decaying asymptotics at the opposite side of the
turning point (its value is of the order $\lambda _{F})$. This is generally
not an easy task but, fortunately, the parameter $z_{0}$ is not important
for finding the boundary conditions for the Green functions.

Substituting Eq.(\ref{wavefunction}) into the spectral expansion, Eq.(\ref
{sum}), we find an expression for the matrix $G^{\alpha \beta }({\bf r};{\bf %
r^{\prime }})$ at the boundary. To determine the matrix $g_{{\bf n}}({\bf r})
$, Eq. (\ref{klassaver}), it is necessary to carry out both the summation in 
$\xi $ (the system may be finite and we have to sum instead of integrating
over $\xi $) and averaging over the energy. The latter is absolutely
necessary because only this averaging guarantees vanishing of all terms
containing the products $ip_{F}{\bf n}\left( {\bf r+r}^{\prime }\right) $ in
the exponents (${\bf n}$ is a unit vector parallel to ${\bf p}$). As soon as
the terms containing $ip_{F}{\bf n}\left( {\bf r+r}^{\prime }\right) $
vanish, the parameter $z_{0}$ drops out. Then, the Green function $G^{\alpha
\beta }({\bf r};{\bf r^{\prime }})$ can be written in the vicinity of the
boundary as

\begin{equation}
G^{\alpha \beta }({\bf r};{\bf r^{\prime }})=i\langle \sum_{k}(\varepsilon
_{k}-\varepsilon )^{-1}  \label{f100}
\end{equation}
\[
\times e^{i{\bf p}_{\shortparallel }\left( {\bf r}_{\shortparallel }-{\bf r}%
_{\shortparallel }^{\prime }\right) }(e^{-ip_{z}\left( z-z^{\prime }\right)
}+e^{+ip_{z}\left( z-z^{\prime }\right) })f_{k}({\bf r}_{\shortparallel },%
{\bf r^{\prime }}_{\shortparallel })\rangle _{\varepsilon } 
\]
where $f_{k}\left( {\bf r}_{\shortparallel }{\bf ,r}_{\shortparallel
}^{\prime }\right) =\varphi _{k}({\bf r}_{\shortparallel })\bar{\varphi}_{k}(%
{\bf r}_{\shortparallel }^{\prime })$ and $\langle ...\rangle _{\varepsilon
} $ stands for averaging over the energy.

Carrying out in Eq. (\ref{f100}) summation over $\xi $ and averaging over
the energy, which is equivalent to integration over $\xi $, and Fourier
transforming with respect to ${\bf r-r}^{\prime }$ we obtain the
quasiclassical function $g_{{\bf n}}\left( {\bf r}\right) $. The function
obtained from Eq. (\ref{f100}) is a slow function of the coordinate ${\bf r+r%
}^{\prime }$ and does not change at the boundary under the replacement ${\bf %
n}_{\perp }\rightarrow -{\bf n}_{\perp }$. Hence, we come to the boundary
condition for the quasiclassical Green functions $g_{{\bf n}}\left( {\bf r}%
\right) $

\begin{eqnarray}
&&g_{{\bf n}_{\perp }}({\bf r})=g_{-{\bf n}_{\perp }}({\bf r}),
\label{condition} \\
&&{\bf r}=(z=0,{\bf r}_{\shortparallel })  \nonumber
\end{eqnarray}
where ${\bf n}_{\perp }$ is the component of the vector ${\bf n}$
perpendicular to the surface. Although Eq. (\ref{condition}) is rather
simple, it has not been written in previous works on the ballistic $\sigma $%
-models.

\widetext

\end{document}